\documentclass[times,review,preprint,authoryear]{elsarticle}

\usepackage{jasr}
\usepackage{framed, multirow}

\usepackage{amssymb}
\usepackage{latexsym}

\usepackage{url}
\usepackage{xcolor}
\definecolor{newcolor}{rgb}{.8,.349,.1}

\usepackage[citebordercolor=white]{hyperref}

\graphicspath{{figures/}}

\journal{Advances in Space Research}

\begin{document}

\verso{Yontan \textit{et al.}}

\begin{frontmatter}

\title{CCD UBV and Gaia DR3 based analysis of NGC 189, NGC 1758 and NGC 7762 open clusters}

\author[1]{Talar \snm{Yontan}\corref{cor1}}
\cortext[cor1]{Corresponding author: 
  Tel.: +90-212-440-0000;  
  fax: +90-212-440-0370;
  Email: talar.yontan@istanbul.edu.tr, 
  \\ sbilir@istanbul.edu.tr, hcakmak@istanbul.edu.tr, rmm@astro.unam.mx, tim.banks@nielsen.com, esoydugan@comu.edu.tr, rmzycnby@gmail.com, tasdemir.seval@ogr.iu.edu.tr}
  
\author[1]{Sel\c{c}uk \snm{Bilir}}
\author[1]{Hikmet \snm{\c{C}akmak}} 
\author[2]{Michel \snm{Ra{\'u}l}}
\author[3,4]{Timothy \snm{Banks}}
\author[5,6]{Esin \snm{Soydugan}}
\author[7]{Remziye \snm{Canbay}}
\author[7]{Seval \snm{Ta\c{s}demir}}

\address[1]{Istanbul University, Faculty of Science, Department of Astronomy and Space Sciences, 34116, Beyaz\i t, Istanbul, Turkey}
\address[2]{Observatorio Astron\'omico Nacional, Universidad Nacional Aut\'onoma de M\'exico, Ensenada, M\'exico}
\address[3]{Department of Physical Science \& Engineering, Harper College, 1200 W Algonquin Rd, Palatine, IL 60067, USA}
\address[4]{Nielsen, 200 W Jackson Blvd, Chicago, IL 60606, USA}
\address[5]{\c{C}anakkale Onsekiz Mart University, Faculty of  Sciences, Department of Physics, 17100,  \c{C}anakkale, Turkey}
\address[6]{\c{C}anakkale Onsekiz Mart University, Astrophysics Research Center and Ulup{\i}nar Observatory, 17100, \c{C}anakkale, Turkey}
\address[7]{Istanbul University, Institute of Graduate Studies in Science, Programme of Astronomy and Space Sciences, 34116, Beyaz{\i}t, Istanbul, Turkey}

\received{10 February 2023}
\finalform{XX XXX 2023}
\accepted{09 April 2023}
\availableonline{XX XXX 2023}
\communicated{T. Yontan}

\begin{abstract}

This paper presents photometric, astrometric, and kinematic analyses of the open clusters NGC 189, NGC 1758 and NGC 7762 based on CCD {\it UBV} photometric and {\it Gaia} Data Release 3 (DR3) data. According to membership analyses, we identified 32, 57 and 106 most probable member stars with membership probabilities $P\geq 0.5$ in NGC 189, NGC 1758 and NGC 7762, respectively. The color excesses and photometric metallicities of each cluster were determined separately using {\it UBV} two-color diagrams. The color excess $E(B-V)$ is $0.590 \pm 0.023$ mag for NGC 189, $0.310 \pm 0.022$ mag for NGC 1758 and $0.640 \pm 0.017$ mag for NGC 7762. The photometric metallicity [Fe/H] is $-0.08 \pm 0.03$ dex for both NGC 189 and NGC 1758, and $-0.12 \pm 0.02$ dex for NGC 7762. Distance moduli and ages of the clusters were obtained by comparing  {\sc PARSEC} isochrones with the color-magnitude diagrams constructed from {\it UBV} and {\it Gaia} photometric data. During this process, we kept as constant color excess and metallicity for each cluster. The estimated isochrone distance is $1201 \pm 53$ pc for NGC 189, $902 \pm 33$ pc for NGC 1758 and $911 \pm 31$ pc for NGC 7762. These are compatible with the values obtained from trigonometric parallax. Ages of the clusters are $500\pm 50$ Myr, $650\pm 50$ Myr and $2000\pm 200$ Myr for NGC 189, NGC 1758 and NGC 7762, respectively. Galactic orbit integration of the clusters showed that  NGC 1758 completely orbits outside the solar circle, while NGC 189 and NGC 7762 enter the solar circle during their orbits.

\end{abstract}


\end{frontmatter}




\section{Introduction}
\label{sec:introduction}

Open clusters consist of stars initially gravitationally bound together, sharing similar positional and kinematic characteristics. Such clusters in our Galaxy are younger and of higher metallicity than the Milky Way's globular clusters. As they formed from the same molecular cloud, member stars share the same general distance, chemical composition and age 
 but vary in formation masses (the initial mass function). These properties make them `laboratories' allowing the study of stellar formation \protect\citep{Scalo_1998, Hopkins_2018}, evolution \protect\citep{Donor_2020}, kinematics \citep{Tarricq_2021}, and dynamics \protect\citep{Krumholz_2019}. Combining results from studies of many open clusters allows them to be used as tracers of the formation and chemical evolution of the Galactic disc \protect\citep{Adamo_2020}. This paper is a part of a wider programme \protect\citep{Bilir_2006a, Bilir_2010, Bilir_2016, Bostanci_2015, Yontan_2015, Yontan_2019, Yontan_2021, Ak_2016, Banks_2020, Akbulut_2021, Koc_2022, Yontan_2023} applying a common methodology, building up such information. It makes use of ground-based photometry in combination with astrometric data from the {\it Gaia} mission \protect\citep{Gaia_DR1, Gaia_DR2, Gaia_EDR3, Gaia_DR3}, allowing the statistical removal of field stars, which act as a contaminant since they are unrelated to the cluster. The precise determination of cluster parameters requires a thorough and careful removal of this contamination, as will be outlined below. 

The {\it Gaia} mission leads the way in the precise determination of cluster membership as well as being helpful for astrometric solutions, trigonometric parallaxes, and radial velocities of open clusters. The third data release of the European Space Agency's {\it Gaia} mission \citep[hereafter {\it Gaia} DR3,][]{Gaia_DR3} contains a great variety of new information, such as an extended radial velocity survey and astrophysical characterisation of {\it Gaia} objects. Due to the longer period data have been collected over, {\it Gaia} DR3 is of higher quality than the earlier {\it Gaia} DR2 \citep{Gaia_DR2} and {\it Gaia} DR1 \citep{Gaia_DR1} releases. As the first part of {\it Gaia} DR3, the third (early) data release {\it Gaia} EDR3 \citep{Gaia_EDR3} contains astrometric, photometric and spectroscopic data for around 1.8 billion objects. {\it Gaia} DR3 has the same source catalogue as {\it Gaia} EDR3 but supplements it in the sense of the large variety of new data products. For many celestial objects, {\it Gaia} DR3 includes their astrophysical parameters estimated from parallaxes, broadband photometry, and mean radial velocity spectra \citep{Gaia_DR3}. In {\it Gaia} DR3, trigonometric parallax errors are 0.02-0.07 mas for $G\leq17$ mag, 0.5 mas for $G=20$ mag and reach 1.3 mas for $G=21$ mag. The proper-motion errors are 0.02~--~0.07 mas yr$^{-1}$, reaching up to 0.5 mas yr$^{-1}$ for $G=20$ mag and 1.4 mas yr$^{-1}$ for $G=21$ mag.

This study explores the clusters NGC 189, NGC 1758 and NGC 7762. These clusters were selected from an ongoing {\em UBVRI} photometric survey of Galactic stellar clusters at the San Pedro Martir Observatory. The current study constitutes the second of our paper series based on this survey \citep{Yontan_2023}, as part of the wider programme described above. The paper outlines an analysis combining CCD {\it UBV} and recent {\it Gaia} DR3 photometric and astrometric data. Independent methods, such as those estimating the color excess and metallicity parameters, reduced possible parameter degeneracy and lead to reliable estimates for astrophysical parameters of all three clusters.


\subsection{NGC 189}

NGC~189 ($\alpha=00^{\rm h} 39^{\rm m} 32^{\rm s}$, $\delta= +61^{\rm o} 05^{\rm '} 24^{\rm''}$, $l = 121\,.\!\!^{\rm o}4938$, $b = -01\,.\!\!^{\rm o}7485$, SIMBAD\footnote{https://simbad.unistra.fr/simbad/} J2000) was likely discovered by Caroline Herschel in 1783 \citep{Hoskin_2006}. The cluster was examined by \cite{Alter_1944} who mentioned that the cluster lies in a comparatively rich region, and estimated its distance as $1490 \pm 200$ pc and diameter $\sim 1.6$ pc. \citet{Balazs_1961} presented color-magnitude and two-color diagrams for the cluster, estimating a distance of 790 pc. \citet{de_la_Fuente_Marcos_2009} noted NGC~189 in their list of candidate binary clusters, matching it with ASCC~4 \citep{Kharchenko_2005}. The cluster has been included in many subsequent studies, such as those based on {\it Gaia} data \citep{Cantat-Gaudin_2018, Liu_2019, Cantat-Gaudin_2020, Dias_2021, Poggio_2021}, with results listed in Table~\ref{tab:literature} for ease of comparison. The table also provides comparisons for NGC 1758 and NGC 7762.


\subsection{NGC 1758}

\cite{Dreyer_1888} noted three overlapping open clusters: NGC 1746, NGC 1750 and NGC 1758 ($\alpha=05^{\rm h} 04^{\rm m} 42^{\rm s}$, $\delta= +23^{\rm o} 48^{\rm '} 47^{\rm''}$, $l = 179\,.\!\!^{\rm o}1584$, $b = -10\,.\!\!^{\rm o}4597$, SIMBAD J2000). Since then there has been disagreement between authors of the existence of one or more of these clusters. \cite{Cuffey_1937}  presented blue and red photographic photometry for NGC~1746 (down to $R \sim 14^{th}$ mag), with no mention of either NGC 1750 or NGC 1758, commenting that NGC 1746 is a thin and asymmetrical cluster. \cite{Straizys_1992} conducted Vilnius photometry \citep{Straivzys_1992} down to {\em V} $\sim$ 13 of 116 stars in the vicinity of the three clusters. They concluded that NGC~1746 is probably not a cluster and were not confident that the other two groupings were real as well. \cite{Tian_1998} concluded there were only two clusters (NGC 1750 and NGC 1758). \cite{Galadi_b_1998, Galadi_1998} disagreed with \citeauthor{Cuffey_1937}'s identification of NGC 1746 based on their Johnson-Cousins CCD {\em UBVRI} photometry, which they judged as completed down to {\em V}~$\sim$ 18.5 mag. \cite{Galadi_1998} wrote that there was no photometric evidence for the cluster NGC~1746 and that \cite{Cuffey_1937} had erroneously assigned a stellar concentration of NGC~1758 as the asymmetric nucleus of NGC~1746. \cite{Galadi_1999} noted NGC 1750 and NGC 1758 to be poor and loose, concluding from their relative velocity, separation, and age difference that they are physically independent clusters. \cite{Landolt_2010} undertook {\em UBV} broadband photometry for 19 stars in the direction of the suspected clusters, admitting confusion on the number of clusters that might be present. Table~\ref{tab:literature} succinctly presents parameters from the literature for NGC 1758, such as color excess, distance moduli, distances, iron abundances, age, proper-motion components, and radial velocity.

\subsection{NGC 7762}

\cite{Chincarini_1966} presented {\em UBV} photographic photometry of NGC~7762 ($\alpha=23^{\rm h} 49^{\rm m} 53^{\rm s}$, $\delta= +68^{\rm o} 02^{\rm '} 06^{\rm''}$, $l = 117\,.\!\!^{\rm o}2060$, $b = 05\,.\!\!^{\rm o}8501$, SIMBAD J2000), classifying the cluster as Trumpler class II 1 m U, 1020 pc away and of diameter 3.5 pc. \cite{Patat_1995} reported CCD {\em BV} photometry of the central region of the cluster, commenting that it was very loose in structure and large.  They concluded that NGC 7762 is of intermediate age, being $\sim 1.8$ Gyr old and $\sim 800$ pc distance. \cite{Szabo_1999} identified photometrically variable stars in the cluster. No $\delta$ Scuti variables were found, which they concluded as confirming that NGC 7762 is older than 800 Myr in age. \cite{Bonatto_2011} presented a proper motion distribution function for the cluster, characterised as a single Gaussian. \cite{Casamiquela_2016} undertook high-resolution spectroscopy of the cluster, providing estimates for $U_{\rm s}$, $V_{\rm s}$ and $W_{\rm s}$ cluster velocities in the Cartesian Galactocentric frame, while \cite{Reddy_2019} analysed high-dispersion Echelle spectra ($R$=60 000) of red giant members to derive a [Fe/H] value of  $-0.07 \pm 0.03$ dex for NGC 7762.

\begin{table*}
\setlength{\tabcolsep}{4pt}
\renewcommand{\arraystretch}{1}
  \centering
  \caption{Fundamental parameters for NGC 189, NGC 1758, and NGC  7762 derived in this study and compiled from the literature: Color excesses ($E(B-V$)), distance moduli ($\mu$), distances ($d$), iron abundances ([Fe/H]), age ($t$), proper-motion components ($\langle\mu_{\alpha}\cos\delta\rangle$, $\langle\mu_{\delta}\rangle$), radial velocity ($V_{\gamma}$) and reference (Ref).}
  \begin{tabular}{ccccccccc}
    \hline
    \hline
    \multicolumn{9}{c}{NGC 189}\\
        \hline
        \hline
$E(B-V)$ & $\mu$ & $d$ & [Fe/H] & $t$ &  $\langle\mu_{\alpha}\cos\delta\rangle$ &  $\langle\mu_{\delta}\rangle$ & $V_{\gamma}$ & Ref \\
(mag) & (mag) & (pc)  & (dex) & (Myr) & (mas yr$^{-1}$) & (mas yr$^{-1}$) & (km s$^{-1})$ &      \\
    \hline
 0.44   & 11.52  & 1080                  & ---    & 2           & ---              & ---              & --- & (01) \\
 0.56   & 11.35  &  860                  & ---    &  ---        & ---              & ---              & --- & (02) \\
 0.700  & ---    & 1300                  & ---    & 510         &$-0.31$           & $-2.71$          & --- & (03)\\
 ---    & ---    & ---                   & ---    & ---         & $-0.36\pm$0.59   & $-3.02\pm$0.10   & --- & (04) \\  
 0.700  & ---    & 1300                  & ---    & 510$\pm$60  & ---              & ---              & --- & (05) \\
 0.420  & ---    &  752                  & ---    & 10          & +0.34$\pm$3.23   & $-1.11 \pm$1.93  & --- & (06) \\
 0.323  & ---    & 1088                  & ---    & 380         & +2.761$\pm$1.368 & $-2.398\pm$0.947 & --- & (07) \\
 ---    & ---    & $1248_{-138}^{+179}$  & ---    & ---         & +0.338$\pm$0.016 & $-3.306\pm$0.023 & --- & (08) \\
 ---    & ---    & $1248_{-138}^{+179}$  & ---    & ---         & +0.338$\pm$0.016 & $-3.306\pm$0.023 & $-28.47\pm$0.75 & (09) \\ 
 ---    & ---    & 1302$\pm$46           & ---    & 562$\pm$34  & +0.323$\pm$0.280 & $-3.307\pm$0.195 & --- & (10) \\
 0.497  & 11.991 & 1228                  & ---    & 400         & +0.338$\pm$0.071 & $-3.306\pm$0.109 & --- & (11) \\ 
 ---    & ---    & $1248_{-138}^{+179}$  & ---    & ---         & +0.338$\pm$0.016 & $-3.306\pm$0.023 & --- & (12) \\ 
 0.602$\pm$0.045 & ---   & 1247$\pm$49           &0.119$\pm$0.152 & 490$\pm$490 & +0.336$\pm$0.120& $-3.315\pm$0.176 & $-29.391\pm$0.534& (13) \\
 0.590$\pm$0.023& 12.227$\pm$0.094 & 1201$\pm$53 & -0.08$\pm$0.03& 500$\pm$50 & +0.352$\pm$0.032  & $-3.412\pm$0.038  & $-29.60\pm$0.32 & (14) \\
  \hline
  \hline
    \multicolumn{9}{c}{NGC 1758}\\
        \hline
        \hline
 0.37$\pm$0.02 & ---   & 680$\pm$24 & ---  &  ---        & ---              & ---              & --- & (15) \\
 0.34$\pm$0.07 & 10.45 & 760        & ---  & 400$\pm$100 &  +0.70           & $-8.1$           & --- & (16) \\
 0.34          & ---   & 760        & ---  & ---         & ---              & ---              & --- & (17) \\
 ---           & ---   & ---        & ---  & ---         & $-0.61\pm$2.32   & $-3.57\pm$0.76   & --- & (04) \\  
 0.34          & ---   & 760        & ---  & 400         & $-0.72\pm$4.83   & $-2.61\pm$3.31   & --- & (06) \\
 0.327         & ---   & 621        & ---  & 540         & $-0.068\pm$0.012 & $+1.335\pm$0.116 & --- & (07) \\
 ---           & ---   & $884_{-72}^{+85}$ & --- & ---   & $+3.156\pm$0.013 & $-3.465\pm$0.010 & --- & (08) \\
 ---           & ---   & 907$\pm$56 & ---  & 355$\pm$21  & $+3.148\pm$0.257 & $-3.474\pm$0.204 & --- & (10) \\
0.299$\pm$0.007& $10.771\pm$0.025   & 931$\pm$11 & 0.00  & $550_{-39}^{+24}$ & ---       & --- & --- & (18) \\
 0.300         & 10.67& 885               & ---    & 355       & $+3.156\pm$0.146 & $-3.465\pm$0.129 & --- & (11) \\ 
 ---           & ---  & $884_{-72}^{+85}$ & ---    & ---       & +3.156$\pm$0.013 & $-3.465\pm$0.010 & --- & (12) \\ 
0.388$\pm$0.019& --- & 864$\pm$29&0.098$\pm$0.073 & 540$\pm$70 & +3.156$\pm$0.157& $-3.470\pm$0.118 & 17.840$\pm$9.887& (13) \\
0.310$\pm$0.022& 10.736$\pm$0.078 & 902$\pm$33 & $-0.08\pm$0.03& 650$\pm$50 & +3.141$\pm$0.035  & $-3.507\pm$0.025  & 6.80$\pm$2.90 & (14) \\
  \hline
  \hline
    \multicolumn{9}{c}{NGC 7762}\\
        \hline
        \hline
 1.02  & 13.10  & 1020                 & ---         &  155 & --- & --- & --- & (19) \\
 0.90  & 12.20  &  800                 & ---         & 1800 & --- & --- & --- & (20) \\
$0.66_{-0.09}^{+0.08}$ & $11.52_{-0.75}^{+0.42}$ & $780_{-300}^{+200}$& --- & 2000 &  --- & --- & --- & (21) \\
 0.59  & $11.3_{-0.8}^{+0.5}$          & $800_{-300}^{+200}$ & 0.00 & $2500_{-700}^{+2000}$ & --- & --- & --- & (22) \\
 0.550 & ---    &  780                 & ---         & 2360$\pm$180 &$-2.88$& 0.00 & --- & (03) \\
 ---   & ---    &  ---                 & ---         & --- & --- & --- &$-45.7\pm$0.7 & (23) \\
 0.550 & ---    & 780                  & ---         & 2100$\pm$100  & --- & ---  & --- & (05) \\
 ---   & ---    & 780 & 0.01$\pm$0.04  &  2500       & --- & --- & --- & (24) \\
 ---   & ---    & $970_{-86}^{+104}$   & ---         & --- & +1.452$\pm$0.010 & $-4.016\pm$0.010 & --- & (08) \\
 ---   & ---    & $970_{-86}^{+104}$   & ---         & --- & +1.452$\pm$0.010 & $-4.016\pm$0.010 & $-45.40\pm$0.14 & (09) \\
 ---   & ---    & --- & $-0.07\pm$0.04 &  ---        & --- & --- & --- & (25) \\
 ---   & ---    & 1005$\pm$41          & ---         & 2630$\pm$158   & +1.478$\pm$0.274 & +4.002$\pm$0.248 & --- & (10) \\
 0.616 & 11.76  & 897                  & ---         & 2050 & +1.452$\pm$0.010 & $- 4.016\pm$0.010 & --- & (11) \\ 
 ---   & ---    & $970_{-86}^{+104}$   & ---         & ---  & +1.452$\pm$0.010 & $-4.016\pm$0.010 & --- & (12) \\ 
 0.840$\pm$0.017& ---                  & 957$\pm$13  & $-0.056\pm$0.103 & 1100$\pm$370 & +1.461$\pm$0.203& +4.006$\pm$0.198 & $-45.438\pm$0.492& (13) \\
 0.640$\pm$0.017& 11.781$\pm$0.072     & 911$\pm$31  & $-0.12\pm$0.02& 2000$\pm$200 & +1.489$\pm$0.023  & +3.962$\pm$0.024  & $-46.61\pm$0.10 & (14) \\
  \hline
    \end{tabular}%
    \\
{\scriptsize
(01) \citet{Lindoff_1968}, (02) \citet{Becker_1971}, (03) \citet{Kharchenko_2013}, (04) \citet{Dias_2014}, (05) \citet{Joshi_2016}, (06) \citet{Sampedro_2017}, (07) \citet{Loktin_2017}, (08) \citet{Cantat-Gaudin_2018}, (09) \citet{Soubiran_2018}, (10) \citet{Liu_2019}, (11) \citet{Cantat-Gaudin_2020}, (12) \citet{Cantat-Gaudin-Anders_2020}, (13) \citet{Dias_2021}, (14) This study, (15) \citet{Straizys_1992}, (16) \citet{Galadi_b_1998}, (17) \citet{Straizys_2003}, (18) \citet{Bossini_2019}, (19) \citet{Chincarini_1966}, (20) \citet{Patat_1995}, (21) \citet{Maciejewski_2007}, (22) \citet{Maciejewski_2008}, (23) \citet{Casamiquela_2016}, (24) \citet{Casamiquela_2017}, (25) \citet{Reddy_2019}
}
\\
  \label{tab:literature}%
\end{table*}%

\section{Observations}
\label{sec:observations}

\subsection{CCD UBV photometric data}
\label{sec:ubv_data}

The observations of these three clusters were collected by R.\ Michel at the San Pedro Martir Observatory as part of an ongoing\footnote{\url{http://bufadora.astrosen.unam.mx/~rmm/SPMO_UBVRI_Survey/Clusters_All.html}} {\em UBVRI} photometric survey of Galactic stellar clusters. In the case of these clusters, the 0.84-m ($f/15$) Ritchey–Chretien telescope was employed along with the Mexman filter wheel and the Marconi 5 camera (a $2048\times2048$ 15$\mu$m square-pixel E2V CCD31-42 detector with a gain of $2.38\mathrm{e^-}$/ADU and a readout noise of $4.02\mathrm{e^-}$ with $2 \times 2$ binning employed, providing an unvignetted field of view of $9.3 \times 9.3$ arcmin$^2$). Exposure times were 2, 20 and 200s for both filters {\it I} and {\it R}; 3, 30 and 300 for filter {\it V}; 5, 50 and 500s for filter {\it B}; and 10, 100 and 1000s for filter {\it U}. 

The observations were carried out during photometric conditions across different observing runs. NGC 189 was observed on 22 September 2015. NGC 1758 was observed on 4 February 2016, and NGC 7762 on 17 January 2015. Landolt's standard stars \citep{Landolt_2009} were also observed, both at the meridian and at around two airmasses, in order to properly determine the transformation equations. Flat fields were taken at the beginning and the end of each night. Bias images were obtained between cluster observations. Data reduction with  point spread function (PSF) photometry was carried out with the IRAF/DAOPHOT packages. The transformation equations recommended by \citet{Stetson19} were employed.

\subsection{Gaia Data}
\label{sec:gaia}

\begin{figure*}
\centering
\includegraphics[scale=1.02, angle=0]{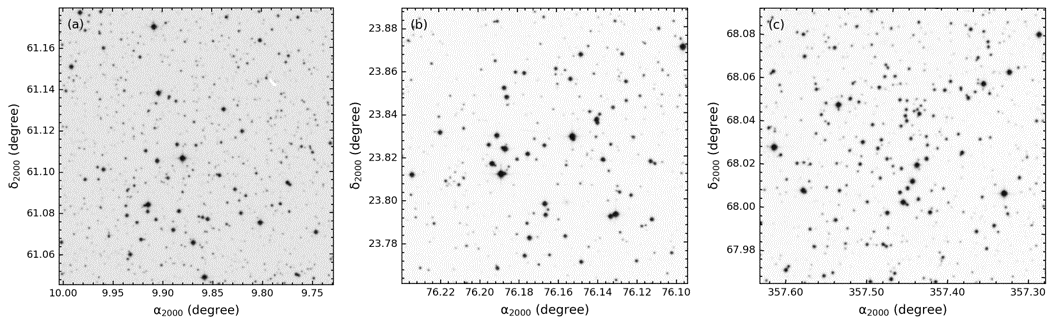}
\caption{Identification charts of NGC 189 (a), NGC 1758 (b) and NGC 7762 (c). The field of view of the charts are $9'.3 \times 9'.3$. North is up and East left.} 
\label{fig:ID_charts}
\end {figure*}

The {\it Gaia} DR3 database \citep{Gaia_DR3} was used together with CCD {\it UBV} photometry to accomplish astrometric, photometric, and kinematic analyses of NGC 189, NGC 1758 and NGC 7762. We extracted data from the {\it Gaia} DR3 database within a 40 arcmin radius (centred on literature cluster centres) and matched it with the {\it UBV} data according to stars' equatorial coordinates, taking into account distances between coordinates of less than 5 arcsec between the data sets. Hence we created photometric and astrometric catalogues for the three clusters containing the detected stars' positions ($\alpha, \delta$), apparent $V$ magnitudes, $U-B$ and $B-V$ color indices, {\it Gaia} DR3 proper-motion components ($\mu_{\alpha}\cos\delta, \mu_{\delta}$), trigonometric parallaxes ($\varpi$), apparent $G$ magnitudes, and $G_{\rm BP}-G_{\rm RP}$ color indices as well as the membership probabilities ($P$) which were calculated in this study (Table ~\ref{tab:all_cat}, page~\pageref{tab:all_cat}). {\it UBV} and {\it Gaia} based photometric and astrometric data for each star located in the cluster areas are available electronically for NGC 189, NGC 1758 and NGC 7762.\footnote{The complete tables can be obtained from VizieR electronically.} $9'.3 \times 9'.3$ field of view optical images for the clusters under study are presented in Figure~\ref{fig:ID_charts}. The photometric errors in {\it UBV} and {\it Gaia} DR3 data were adopted as internal errors, which are the uncertainties of the instrumental magnitudes of the stars. The mean errors in the {\it UBV} and {\it Gaia} DR3 photometric data were calculated as functions of $V$ apparent magnitude. The results are given in Table~\ref{tab:photometric_errors} (on page~\pageref{tab:photometric_errors}). We found that the mean errors in $V$ magnitude reach up to 0.1 mag, while for the $U-B$ and $B-V$ color indices, the mean errors are about 0.3 mag and 0.2 mag, respectively, for the three clusters. The mean errors in $G$ magnitudes reach up to 0.01 mag, whereas in the $G_{\rm BP}-G_{\rm RP}$ color index they are less than 0.25 mag for the stars $V\leq22$ mag (see Table~\ref{tab:photometric_errors}).  

\begin{figure*}
\centering
\includegraphics[scale=0.6, angle=0]{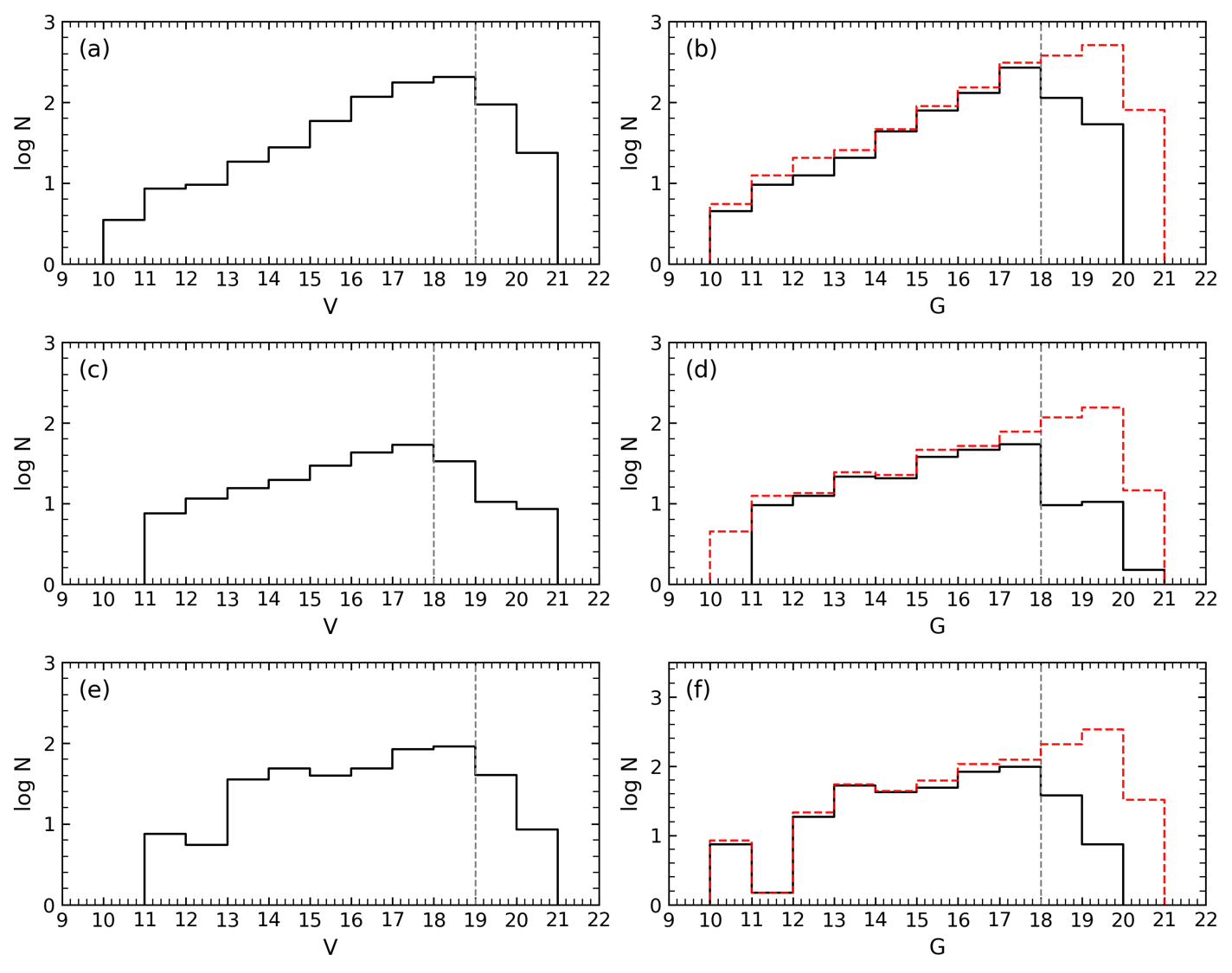}\\
\caption{Histograms of NGC 189 (a, b), NGC 1758 (c, d) and NGC 7762 (e, f) per magnitude bin in the filters $V$ (black lines) and $G$ (red lines). The vertical dashed lines indicate the adopted faint limiting apparent magnitudes in the $V$ and $G$ bands.} 
\label{fig:histograms}
\end {figure*}

Crowding by other stellar images in CCD frames can prevent the detection of faint stars in the clusters. This crowding will lower the number of detected faint stars, dropping the `completeness' from unity (where all stars are detected). Understanding the decreasing completeness of the measured stellar counts with increasing magnitude is necessary to derive reliable astrophysical parameters of open clusters. To define photometric completeness limits for the three clusters, we constructed histograms of the observed $G$ and $V$ magnitudes (Fig.~\ref{fig:histograms}). We compared the observed $G$ magnitude histograms with the {\it Gaia} DR3 data for the $9'\!.3 \times 9'\!.3$ region on the sky matching the fields observed from the ground. All histograms for stellar counts are shown in Fig.~\ref{fig:histograms}: the black solid lines show the observational stellar distributions by magnitude bin in $V$ and $G$, while the red solid lines (see in Fig.~\ref{fig:histograms}b, \ref{fig:histograms}d and \ref{fig:histograms}f) represent the {\it Gaia} DR3 based stellar counts. It can be seen in Fig.~\ref{fig:histograms}b, \ref{fig:histograms}d and \ref{fig:histograms}f that the numbers of stars detected in the three cluster regions are well matched with the {\it Gaia} DR3 stellar distributions up to fainter magnitudes. The magnitudes adopted as photometric completeness limits are $V=19$ mag for NGC 189 and NGC 7762, $V=18$ mag for NGC 1758 and $G=18$ mag for all clusters (the limits are the vertical dashed lines in all panels of Fig.~\ref{fig:histograms}). In Fig.~\ref{fig:histograms}b, \ref{fig:histograms}d and \ref{fig:histograms}f the number of stars decreases with the increasing crowding of low-mass stars beyond the adopted completeness limits. Observational techniques, telescope-detector combinations, and telescope qualities used in ground and space-based observations have an impact on detecting stars, especially at fainter magnitudes. This could clarify the reason why the stellar counts fainter than $G>18$ mag in the {\it Gaia} space-based observations are greater than the number of detected stars in the study.   

\begin{table*}
\setlength{\tabcolsep}{3.5pt}
\renewcommand{\arraystretch}{1.05}
\scriptsize
  \centering
  \caption{\label{tab:input_parameters}
{The photometric and astrometric catalogues for NGC 189, NGC 1758, and NGC 7762 members}}.
    \begin{tabular}{cccccccccccc}
\hline
\hline
\multicolumn{11}{c}{NGC 189}\\
\hline
\hline
ID	 & RA           &	DEC	        &      $V$	    &	$U-B$      & $B-V$	      &	$G$	          & 	$G_{\rm BP}-G_{\rm RP}$	 & 	$\mu_{\alpha}\cos\delta$ & 	$\mu_{\delta}$ & 	$\varpi$	& $P$ \\
	 & (hh:mm:ss.ss) &	(dd:mm:ss.ss)	&      (mag)	    &	(mag)      & (mag)	      &	(mag)	          & 	(mag)	 & 	(mas yr$^{-1}$) & 	(mas yr$^{-1}$) & 	(mas)	&  \\
\hline
\hline
001 & 00:39:31.15 & +61:06:23.67 & 10.933(0.008) & 0.522(0.065) & 0.863(0.064) & 10.844(0.011) & 1.302(0.005) & ---           & ---           & ---          & --- \\
002 & 00:39:37.97 & +61:10:12.88 & 11.221(0.025) & 0.202(0.045) & 0.590(0.050) & 11.210(0.009) & 0.854(0.005) & ---           & ---           & ---          & --- \\
003 & 00:39:28.64 & +61:03:56.79 & 11.831(0.029) & 0.331(0.040) & 0.443(0.047) & 11.789(0.003) & 0.611(0.005) & -6.112(0.012) & 0.382(0.014)  & 1.097(0.013) & 0.00 \\
004 & 00:39:12.49 & +61:04:31.15 & 11.837(0.019) & 0.428(0.038) & 0.628(0.041) & 11.655(0.003) & 0.935(0.005) & -0.628(0.037) & -3.088(0.043) & 0.859(0.39)  & 0.00 \\
... & ... & ...	& ... & ... & ... & ... & ... & ... & ... & ... & ...\\
... & ... & ...	& ... & ... & ... & ... & ... & ... & ... & ... & ...\\
733 & 00:39:45.99 & +61:06:59.69 & 21.435(0.179) & ---          & 0.945(0.234) & 20.808(0.012) & 1.424(0.210) & -2.461(01.585) & 1.607(1.503) & 0.727(1.303) & 0.00 \\ 
734 & 00:39:13.34 & +61:07:07.29 & 21.583(0.165) & ---          & 0.673(0.211) & ---           & ---          & ---            & ---          &              & --- \\ 
735 & 00:39:04.22 & +61:05:07.37 & 21.587(0.187) & ---          & 1.428(0.333) & 20.924(0.022) & 1.843(0.368) & ---            & ---          &              & --- \\ 
736 & 00:39:25.31 & +61:05:25.71 & 21.655(0.272) & ---          & 0.886(0.317) & 20.744(0.010) & 1.340(0.176) & -2.047(1.038 ) & 1.115(1.132) & 0.025(0.925) & 0.01 \\ 
\hline
\hline
\multicolumn{11}{c}{NGC 1758}\\
\hline
\hline
001 & 05:04:22.95 & +23:52:14.48 & 11.105(0.074) & 0.267(0.101) & 0.429(0.111) & 11.009(0.003) & 0.641(0.005) & 3.112(0.022) & -3.649(0.016) & 1.179(0.020) & 0.99 \\
002 & 05:04:31.31 & +23:47:34.80 & 11.346(0.071) & 0.254(0.092) & 0.413(0.103) & 11.266(0.003) & 0.620(0.005) & 3.019(0.025) & -3.438(0.019) & 1.117(0.021) & 1.00 \\
003 & 05:04:46.38 & +23:49:00.94 & 12.101(0.048) & 0.334(0.068) & 0.434(0.075) & 11.997(0.003) & 0.662(0.005) & 3.136(0.018) & -3.547(0.013) & 1.138(0.016) & 1.00 \\ 
004 & 05:04:56.17 & +23:48:43.25 & 12.334(0.045) & 0.322(0.062) & 0.401(0.070) & 12.200(0.003) & 0.633(0.005) & 2.927(0.020) & -3.330(0.014) & 1.159(0.017) & 0.99 \\
... & ... & ...	& ... & ... & ... & ... & ... & ... & ... & ... & ...\\
... & ... & ...	& ... & ... & ... & ... & ... & ... & ... & ... & ...\\
226 & 05:04:26.84 & +23:46:42.32 & 21.308(0.127) & ---          & 0.882(0.193) & 21.004(0.025) & 0.854(0.444) & ---           & ---            &  ---          & 0.47 \\ 
227 & 05:04:40.16 & +23:51:46.41 & 21.359(0.132) & ---          & 0.253(0.176) & ---		   & --- 		  & ---           & ---            &  ---          & --- \\ 
228 & 05:04:37.57 & +23:48:04.10 & 21.385(0.162) & ---          & 0.720(0.217) & 20.724(0.012) & 0.641(0.260) & 0.349 (2.355) & -0.566(1.632) & -1.666(1.586) & 0.00 \\ 
229 & 05:04:46.99 & +23:47:08.86 & 21.585(0.170) & ---          & 0.939(0.260) & 20.696(0.011) & 0.968(0.427) & -0.428(1.428) & -1.276(1.079) & 0.279 (1.264) & 0.00 \\ 
\hline
\hline
\multicolumn{11}{c}{NGC 7762}\\
\hline
\hline
001 & 23:49:21.67 & +68:01:01.74 & 11.705(0.011) & 1.218(0.012) & 1.700(0.013) & 10.794(0.003) & 2.131(0.005) & 1.290(0.018) & 3.747 (0.019) & 1.034(0.017) & 0.98 \\ 
002 & 23:50:31.42 & +68:01:41.56 & 11.725(0.017) & 1.204(0.014) & 1.609(0.019) & 10.904(0.003) & 2.010(0.005) & 1.345(0.021) & 4.134 (0.022) & 1.017(0.019) & 0.98 \\ 
003 & 23:50:13.51 & +68:03:02.70 & 12.199(0.012) & 1.094(0.009) & 1.559(0.013) & 11.482(0.003) & 1.925(0.005) & 3.135(0.015) & -3.269(0.016) & 1.034(0.014) & 0.00 \\
004 & 23:49:23.53 & +68:04:24.88 & 12.471(0.052) & 0.726(0.023) & 1.266(0.056) & 11.835(0.003) & 1.763(0.005) & 3.358(0.177) & 0.207 (0.182) & 1.661(0.174) & 0.00 \\ 
... & ... & ...	& ... & ... & ... & ... & ... & ... & ... & ... & ...\\
... & ... & ...	& ... & ... & ... & ... & ... & ... & ... & ... & ...\\
403 & 23:50:01.51 & +68:01:21.45 & 21.300(0.146) &---           & 0.457(0.190) & ---		   & --- 		  & ---           & ---           &  ---         & --- \\ 
404 & 23:49:28.77 & +68:01:48.92 & 21.390(0.178) &---           & 1.349(0.296) & 20.462(0.007) & 2.040(0.138) & -0.890(0.759) & -1.817(0.634) & -0.506(0.581)& 0.00 \\ 
405 & 23:50:11.55 & +68:01:49.47 & 21.491(0.205) &---           & 1.227(0.326) & --- & ---		   & --- 		  & ---           & ---           &  ---\\ 
406 & 23:49:34.88 & +67:58:37.81 & 21.912(0.267) &---           & 0.575(0.355) & 20.838(0.011) & 1.772(0.235) & -1.741(1.444) & 1.070 (1.400) & -1.654(1.290)& 0.00 \\ 
\hline
\hline
    \end{tabular}
      \label{tab:all_cat}%
\end{table*} 

\begin{table}
\setlength{\tabcolsep}{9pt}
\renewcommand{\arraystretch}{1.2}
  \centering
  \caption{Mean internal photometric errors per magnitude bin in $V$ brightness for the three open clusters.}
    \begin{tabular}{ccccccc}
      \hline
      \\[-0.5em]
    \multicolumn{7}{c}{NGC 189}\\
    \hline
  $V$ & $N$ & $\sigma_{\rm V}$ & $\sigma_{\rm U-B}$ & $\sigma_{\rm B-V}$ & $\sigma_{\rm G}$ &  $\sigma_{G_{\rm BP}-G_{\rm RP}}$\\
  \hline
  (10, 12] &   4 & 0.020 & 0.047 & 0.050 & 0.006 & 0.005 \\
  (12, 14] &  17 & 0.015 & 0.026 & 0.028 & 0.003 & 0.005 \\
  (14, 15] &  18 & 0.009 & 0.015 & 0.014 & 0.003 & 0.005 \\
  (15, 16] &  27 & 0.009 & 0.018 & 0.012 & 0.003 & 0.007 \\
  (16, 17] &  58 & 0.011 & 0.024 & 0.014 & 0.003 & 0.006 \\
  (17, 18] & 117 & 0.015 & 0.041 & 0.021 & 0.003 & 0.010 \\
  (18, 19] & 175 & 0.025 & 0.089 & 0.037 & 0.003 & 0.021 \\
  (19, 20] & 204 & 0.038 & 0.171 & 0.062 & 0.003 & 0.039 \\
  (20, 22] & 116 & 0.103 & 0.295 & 0.160 & 0.005 & 0.100 \\
   \hline
   \\[-0.5em]
          \multicolumn{7}{c}{NGC 1758}\\
    \hline
  $V$ & $N$ & $\sigma_{\rm V}$ & $\sigma_{\rm U-B}$ & $\sigma_{\rm B-V}$ & $\sigma_{\rm G}$ &  $\sigma_{G_{\rm BP}-G_{\rm RP}}$\\
  \hline
  (10, 12] &   4 & 0.072 & 0.097 & 0.107 & 0.003 & 0.005 \\
  (12, 14] &  18 & 0.020 & 0.042 & 0.044 & 0.003 & 0.005 \\
  (14, 15] &  15 & 0.010 & 0.022 & 0.021 & 0.003 & 0.005 \\
  (15, 16] &  19 & 0.006 & 0.013 & 0.009 & 0.003 & 0.006 \\
  (16, 17] &  29 & 0.011 & 0.017 & 0.014 & 0.003 & 0.008 \\
  (17, 18] &  42 & 0.012 & 0.040 & 0.020 & 0.003 & 0.014 \\
  (18, 19] &  53 & 0.021 & 0.074 & 0.032 & 0.003 & 0.030 \\
  (19, 20] &  32 & 0.036 & 0.082 & 0.073 & 0.004 & 0.065 \\
  (20, 22] &  17 & 0.099 & 0.226 & 0.158 & 0.010 & 0.240 \\
  \hline
   \\[-0.5em]
            \multicolumn{7}{c}{NGC 7762}\\
    \hline
  $V$ & $N$ & $\sigma_{\rm V}$ & $\sigma_{\rm U-B}$ & $\sigma_{\rm B-V}$ & $\sigma_{\rm G}$ &  $\sigma_{G_{\rm BP}-G_{\rm RP}}$\\
  \hline
  (10, 12] &   2 & 0.014 & 0.013 & 0.016 & 0.003 & 0.005 \\
  (12, 14] &  12 & 0.019 & 0.013 & 0.021 & 0.003 & 0.005 \\
  (14, 15] &  35 & 0.014 & 0.013 & 0.016 & 0.003 & 0.005 \\
  (15, 16] &  48 & 0.008 & 0.016 & 0.010 & 0.003 & 0.006 \\
  (16, 17] &  39 & 0.010 & 0.031 & 0.014 & 0.003 & 0.006 \\
  (17, 18] &  48 & 0.014 & 0.077 & 0.026 & 0.003 & 0.011 \\
  (18, 19] &  84 & 0.022 & 0.133 & 0.043 & 0.003 & 0.022 \\
  (19, 20] &  90 & 0.038 & 0.186 & 0.076 & 0.003 & 0.048 \\
  (20, 22] &  48 & 0.099 & ---   & 0.184 & 0.005 & 0.100 \\
  \hline
    \end{tabular}%
  \label{tab:photometric_errors}%
\end{table}%

\subsection{Spatial Structures of the Clusters}
\label{sec:structural}
Before investigating the spatial structure of the three clusters, we determined the central equatorial coordinates ($\alpha$, $\delta$) by applying the star count method. The calculated central coordinates for NGC 189 and NGC 7762 are in agreement with the values given by \citet{Cantat-Gaudin_2020}. Results are listed in Table~\ref{tab:rdp} (on page \pageref{tab:rdp}) for the three clusters. 

\begin{figure}[t]
\centering
\includegraphics[scale=0.23, angle=0]{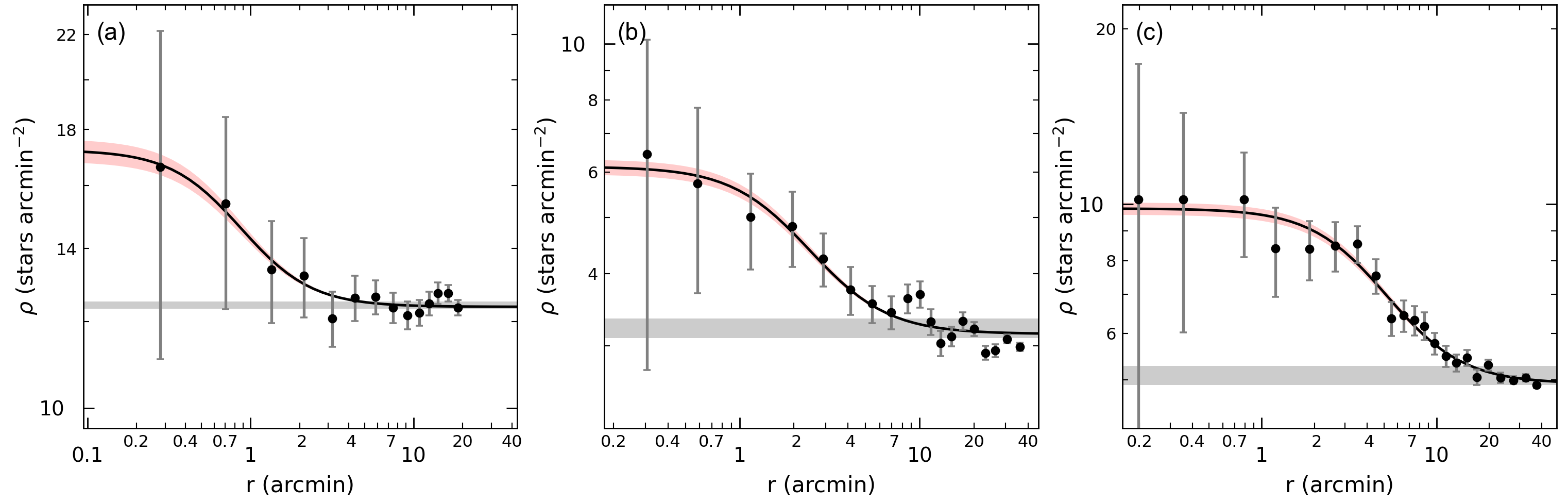}\\
\caption{The stellar density distribution of the clusters NGC 189 (a), NGC 1758 (b), and NGC 7762 (c). The fitted black curve shows the RDP profile of \citet{King_1962}, whereas the horizontal grey band represents the background stellar density. The $1\sigma$ King fit uncertainty is pictured by the red-shaded area.} 
\label{fig:king}
\end {figure}

To measure the extent of the clusters,  we used {\it Gaia} DR3 data within 40 arcmin of the derived cluster centres. Estimation of the  structural parameters of NGC 189, NGC 1758 and NGC 7762 employed a radial density profile (RDP) for each cluster as shown in Figure~\ref{fig:king} (page~\pageref{fig:king}). We divided a cluster's area into several concentric rings around the cluster center that we estimated in the study and calculated the stellar number density ($\rho(r)$) for an $\it i$th ring as $R_{i}=N_{i}/A_{i}$, where $N_{i}$ and $A_{i}$ indicate the number of stars in that circle and the surface area, respectively. Uncertainties in the stellar number densities were estimated by Poisson statistics ($1/\sqrt N$, where $N$ is the number of stars).  We fitted the empirical RDP of \citet{King_1962}, which is expressed by the following formula:

\begin{equation}
\rho(r)=f_{\rm bg}+\frac{f_0}{1+(r/r_{\rm c})^2}
\end{equation}
where $f_{\rm bg}$ is the background stellar density, $f_0$ the central stellar density and $r_{\rm c}$ the core radius of the cluster. The RDP fitting procedure utilized a $\chi^{2}$ minimization method. The derived structural parameters for each cluster are listed in Table~\ref{tab:rdp} (page~\pageref{tab:rdp}). The best-fit RDP of each cluster is represented by a solid line in the corresponding sub-figures of Figure~\ref{fig:king}. We estimated the limiting radii ($r_{\rm lim}^{\rm obs}$) according to a visual review of RDP. We identified that the limiting radius as the point where the background stellar density (shown by the horizontal grey band in Figure~\ref{fig:king}) matches with or equals the RDP. We therefore considered the limiting radii as $r_{\rm lim}^{\rm obs}=4'$ for NGC 189, $r_{\rm lim}^{\rm obs}=7'$ for NGC 1758 and $r_{\rm lim}^{\rm obs}=15'$ for NGC 7762. To determine the mathematical accuracy of these visually determined $r_{\rm lim}^{\rm obs}$ values, we used the equation of \citet{Bukowiecki_2011}:

\begin{equation}
r_{\rm lim}=r_{\rm c}\sqrt{\frac{f_0}{3\sigma_{\rm bg}}-1}
\end{equation}
We therefore calculated the limiting radii ($r_{\rm lim}^{\rm cal}$) as $3'.55$ for NGC 189, $7'.18$ for NGC 1758 and $14'.95$ for NGC 7762 (see also Table~\ref{tab:rdp}). The fact that the calculated limiting radii are compatible with the observational values, and that the correlation coefficients of the King models with the best-fit model parameters to the data are higher than $R^2=0.95$, indicates that the estimated structural parameters are well estimated.

\newcommand{\fscale}{(stars arcmin$^{-2}$)}
\begin{table}
\setlength{\tabcolsep}{3pt}
\renewcommand{\arraystretch}{1.2}
  \centering
  \caption{The structural parameters of the three open clusters according to \citet{King_1962} model analyses: $(\alpha, \delta)$, $f_0$, $f_{\rm bg}$, $r_{\rm c}$, $r_{\rm lim}^{\rm obs}$, $r_{\rm lim}^{\rm cal}$ and $R^2$ are equatorial coordinates, central stellar densities, the background stellar densities, the core radius, observational limiting radius, calculated limiting radius and correlation coefficients, respectively.}
    \begin{tabular}{lcccccccc}
\hline
  Cluster & $\alpha_{\rm J2000}$ & $\delta_{\rm J2000}$  & $f_0$   & $f_{\rm bg}$ & $r_{\rm c}$ & $r_{\rm lim}^{\rm obs}$ & $r_{\rm lim}^{\rm cal}$& $R^2$\\ 
          & (hh:mm:ss.ss)       &   (dd:mm:ss.ss)        & \fscale &  \fscale     &  (arcmin)   &  (arcmin) & (arcmin)   & \\
\hline
NGC ~~189 & 00:39:27.89 & +61:06:46.35 & 4.86$\pm$0.41 & 12.38$\pm$0.08 & 0.81$\pm$0.12 & ~~4  & ~~3.55 & 0.959\\
NGC 1758  & 05:04:42.00 & +23:48:46.90 & 2.97$\pm$0.17 & ~3.14$\pm$0.08 & 2.13$\pm$0.28 & ~~7  & ~~7.18 & 0.951\\
NGC 7762  & 23:49:53.00 & +68:02:06.00 & 4.91$\pm$0.24 & ~4.92$\pm$0.13 & 4.39$\pm$0.46 &  15  &  14.95 & 0.957\\
\hline
    \end{tabular}%
    \label{tab:rdp}%
\end{table}%

\subsection{Color-Magnitude diagrams and membership probabilities of stars}
\label{sec:cmds}
Open clusters are located through the Galactic plane, leading to them usually being affected by field star contamination. To obtain fundamental cluster parameters more precisely, it is important to separate cluster members from foreground/background non-member stars. Stars in an open cluster are formed under the same physical conditions, and they move in the same general vectorial directions in space. These properties make proper-motion components useful tools to separate cluster members \citep{Bisht_2020} from field stars. Thanks to {\it Gaia} DR3's high accuracy astrometric data, membership analyses give reliable results if carefully made. To determine membership probabilities ($P$) of stars located in the direction of the three clusters, we performed the Unsupervised Photometric Membership Assignment in Stellar Clusters \citep[{\sc upmask};][]{Krone-Martins_2014} method on {\it Gaia} DR3 astrometric data. The method was used in previous studies by various researchers \citep{Cantat-Gaudin_2018, Cantat-Gaudin_2020, Castro-Ginard_2018, Castro-Ginard_2019, Castro-Ginard_2020, Banks_2020, Akbulut_2021, Wang_2022, Yontan_2023}. {\sc upmask} is a $k$-means clustering method which considers the proper-motion components as well as the trigonometric parallaxes of the stars. This technique allows the detection of similar groups of stars, together with estimates of membership probabilities for these groups. To determine the most probable members for NGC 189, NGC 1758 and NGC 7762 and to evaluate membership probabilities, we ran 100 iterations of {\sc upmask} which considered each detected star's astrometric measurements ($\alpha$, $\delta$, $\mu_{\alpha}\cos \delta$, $\mu_{\delta}$, $\varpi$) and their uncertainties. We determined 79 likely member stars for NGC 189, 61 for NGC 1758 and 118 stars for NGC 7762 as the most probable cluster members with membership probabilities $P\geq0.5$ and brighter than the photometric completeness limits described above. 

\begin{figure}[t]
\centering
\includegraphics[scale=0.93, angle=0]{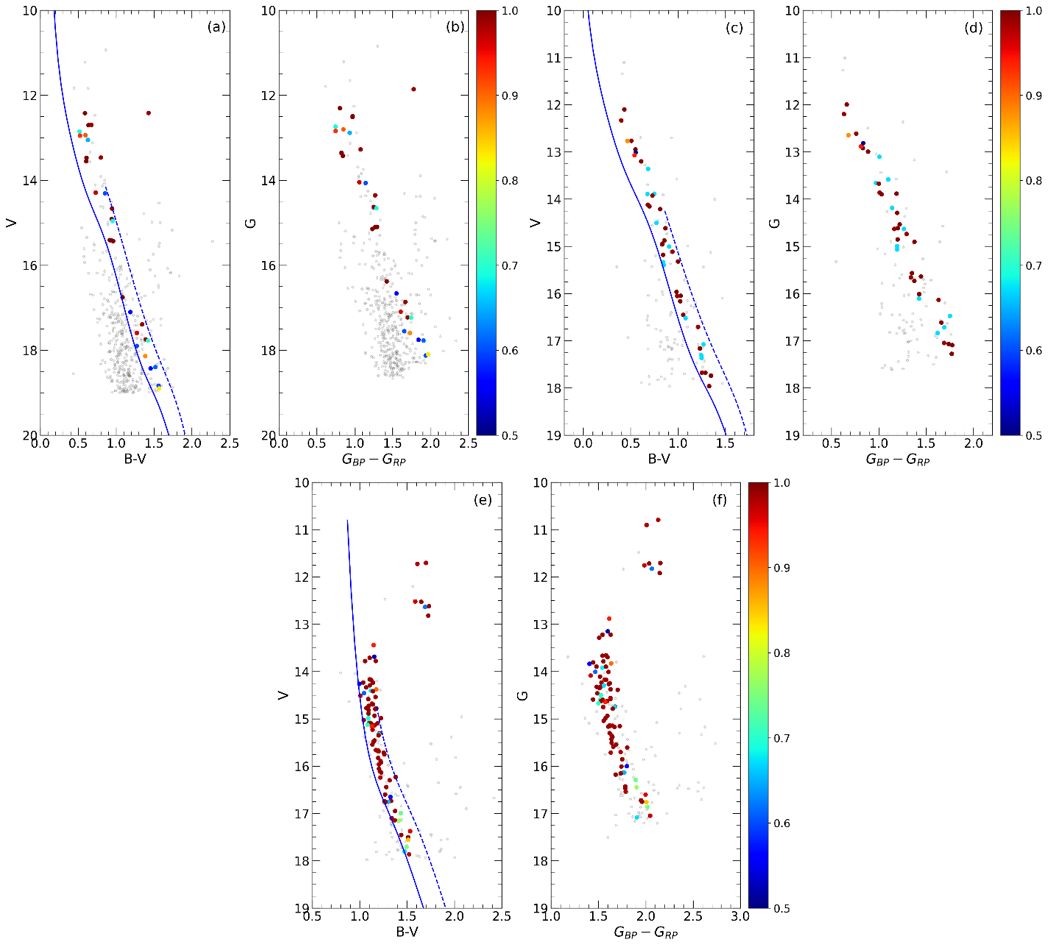}\\
\caption{CMDs of NGC 189 (a, b), NGC 1758 (c, d), and NGC 7762 (e, f) based on {\it UBV} (a, c, e) and {\it Gaia} DR3 photometry (b, d, f). The blue dashed lines show the ZAMS \citep{Sung_2013} including the binary star contamination. The membership probabilities of stars are represented with different colors, the member stars are located within $r_{\rm lim}=4'$, $r_{\rm lim}=7'$, and $r_{\rm lim}=15'$ of the cluster centres calculated for NGC 189, NGC 1758 and NGC 7762, respectively. The stars with low membership probabilities ($P<0.5$) were indicated with grey dots.} 
\label{fig:cmds}
\end {figure}

The observed color-magnitude diagrams (CMDs) are useful tools to derive age, distance, and other key parameters for open star clusters. In other respects, CMDs created in different photometric systems are effective tools to explore the main sequence, turn-off, and giant region of star clusters. In this study, we used {\em UBV} based CMDs to take into account the possible `contamination' by binary stars in the main sequence of the three clusters. Considering the stars with the probabilities $P\geq0.5$, we constructed $V\times (B-V)$ CMDs and fitted the Zero Age Main-Sequence (ZAMS) line of \citet{Sung_2013} to these diagrams through visual inspection according to most probable main-sequence stars. We shifted the ZAMS by 0.75 mag over the brighter stars to adjust for possible binary star contamination by making sure that we chose clearly the most probable main sequence, turn-off and giant members. Consequently, with constraints on the completeness-estimated $V$ magnitude limits and fitted ZAMS, we obtained 32 member stars for NGC 189, 57 for NGC 1758 and 106 for NGC 7762 with membership probabilities $P\geq0.5$. The $V\times (B-V)$ diagrams with fitted ZAMS and distribution of both field and cluster stars are shown in Figs.~\ref{fig:cmds}a (for NGC 189), \ref{fig:cmds}c (for NGC 1758) and \ref{fig:cmds}e (for NGC 7762). The stellar distributions using the {\it Gaia} based photometry are shown in Figs.~\ref{fig:cmds}b (for NGC 189), \ref{fig:cmds}d (for NGC 1758) and \ref{fig:cmds}f (for NGC 7762).

\begin{figure*}[!t]
 \centering
 \includegraphics[scale=0.8, angle=0]{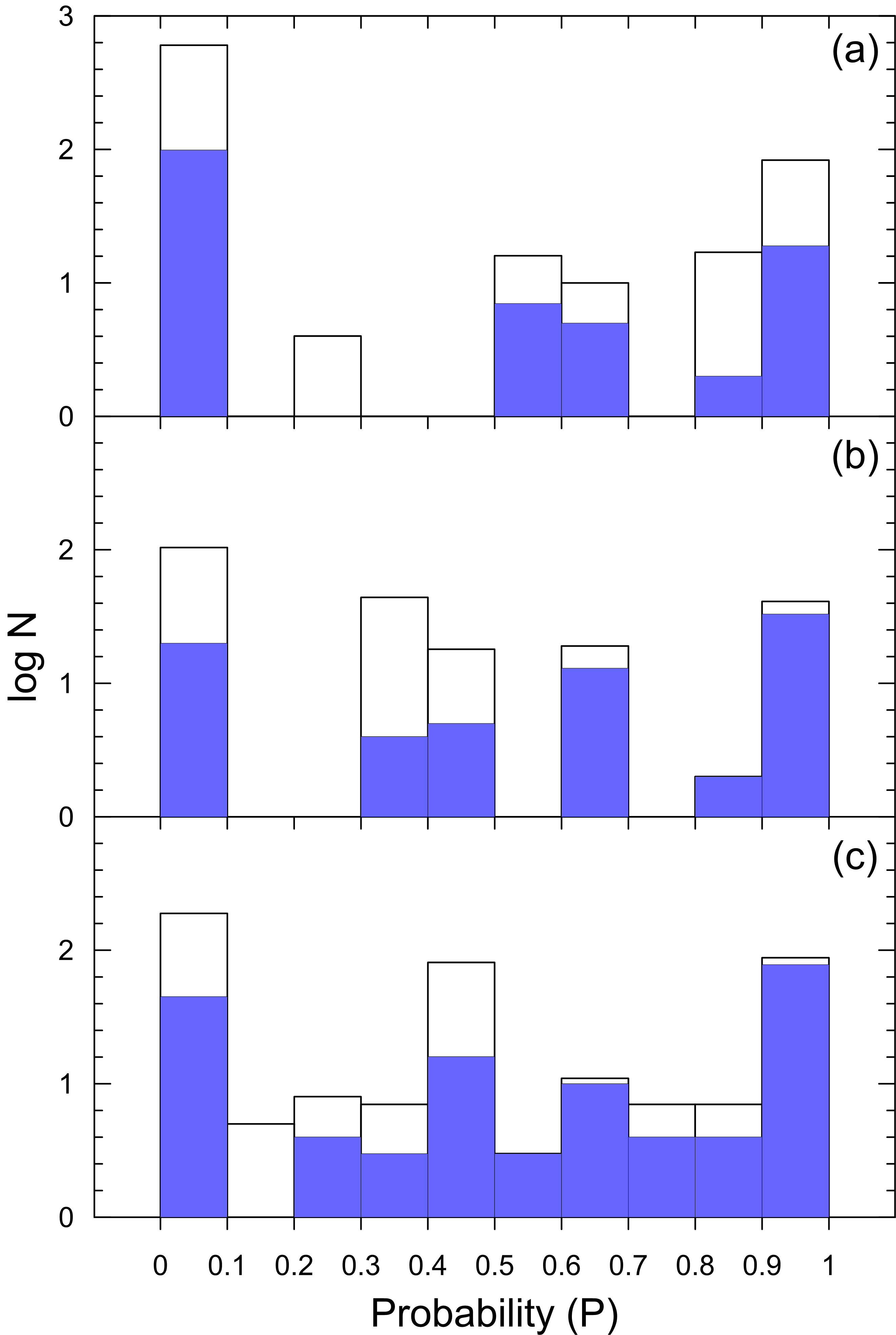}
 \caption{Histogram of the membership probabilities for the stars in the NGC 189 (a), NGC 1758 (b), and NGC 7762 (c) open clusters. The white shading presents the stars that are detected in the cluster areas, while the purple colored shading denotes the stars that lie within fitted ZAMS lines.
 \label{fig:prob_hists} }
 \end {figure*}

\begin{figure}
\centering
\includegraphics[scale=0.70, angle=0]{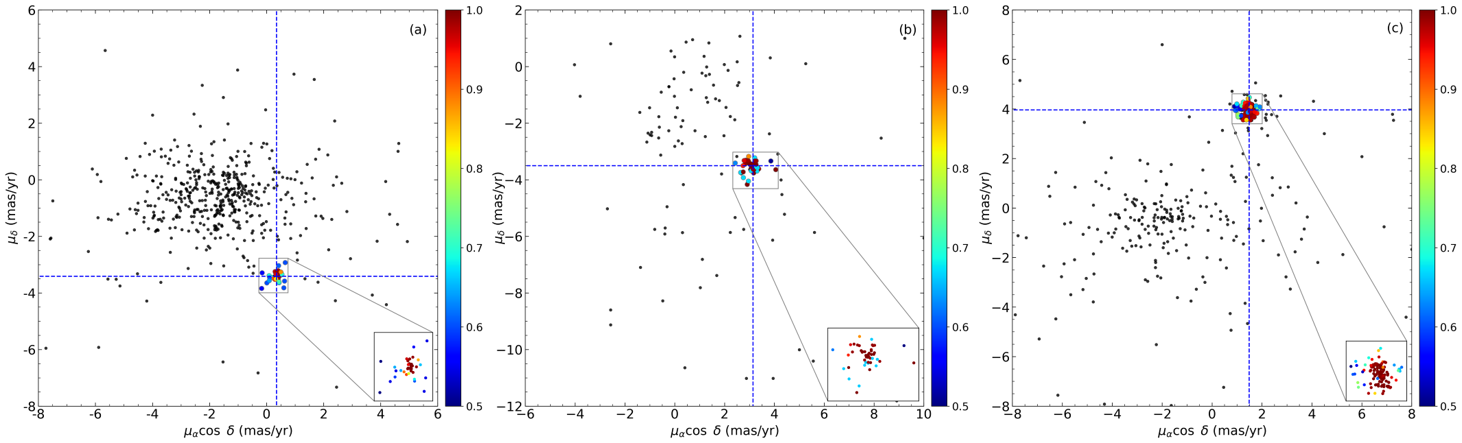}\\
\caption{VPDs of NGC 189 (a), NGC 1758 (b), and NGC 7762 (c) based on {\it Gaia} DR3 astrometry. The membership probabilities of the stars are identified with the color scale shown on the right for each cluster. The zoomed boxes in panels (a), (b) and (c) represent the region of condensation for each cluster in the VPDs. The mean proper motion value of the clusters is shown with the intersection of the dashed blue lines.
\label{fig:VPD_all}} 
 \end {figure}

\begin{figure*}[!t]
\centering
\includegraphics[scale=.74, angle=0]{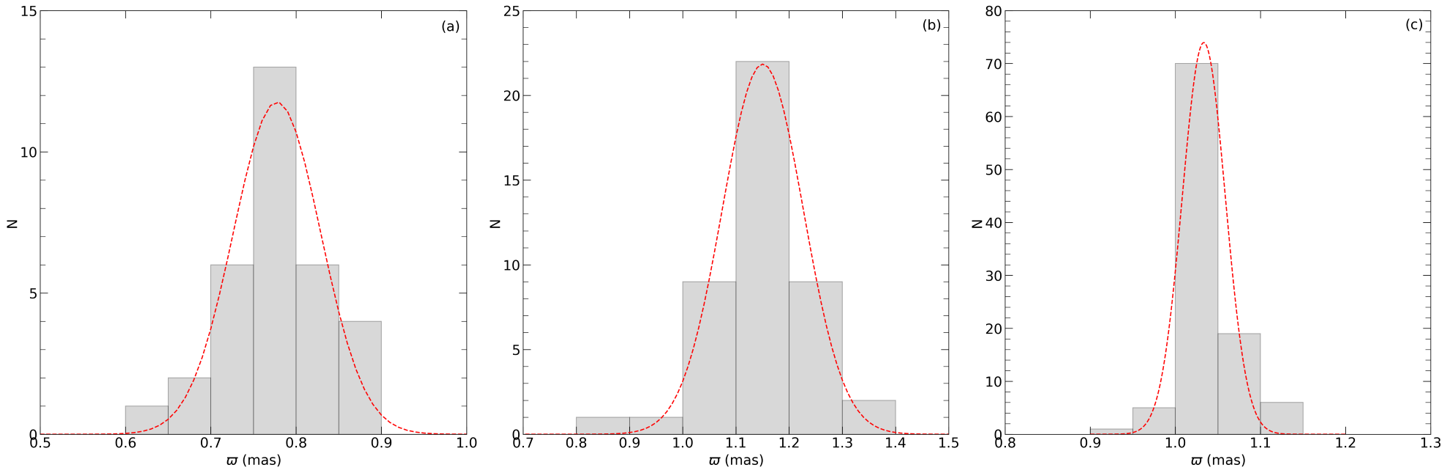}\\
\caption{Histograms of mean trigonometric parallax estimation of NGC 189 (a), NGC 1758 (b), and NGC 7762 (c) from the most probable members ($P\geq 0.50$). Red dashed lines show the fitted Gaussian functions.
\label{fig:plx_hist}}
\end {figure*}

Figure~\ref{fig:prob_hists} plots the membership probabilities of detected stars as a function of stellar numbers. To determine the locations of the most probable cluster members and the field stars in proper-motion space, we drew vector point diagrams (VPDs) for the three clusters (see Fig.~\ref{fig:VPD_all}). It can be seen from the figure that all three clusters are clearly separated from their field stars. In Fig.~\ref{fig:VPD_all}, the intersection of the blue dashed lines indicates the mean proper-motion values estimated from the stars with membership probabilities $P\geq0.5$. The mean proper-motion components are listed in Table~\ref{tab:proper_motion}. These results are in good agreement with the values of previous studies based on {\it Gaia} observations for the three clusters (see Table~\ref{tab:literature}). To estimate the astrometric distance of each cluster, we used trigonometric parallaxes considering the cluster members (i.e., stars with $P\geq0.5$). We constructed the trigonometric parallax histograms using these members and calculated the mean values for each cluster by fitting Gaussian functions to the distributions as shown in Fig.~\ref{fig:plx_hist}. As a result, the mean trigonometric parallax values were obtained as $\varpi=0.778\pm 0.051$ mas for NGC 189, $\varpi=1.150\pm 0.076$ mas for NGC 1758 and $\varpi=1.033\pm 0.025$ mas for NGC 7762. By applying the linear equation of $d({\rm pc})=1000/\varpi$ (mas) to the mean trigonometric parallaxes, we derived the astrometric distances as $d_{\varpi}=1285\pm 84$ pc for NGC 189, $d_{\varpi}=870\pm 57$ pc for NGC 1758 and $d_{\varpi}=968\pm 23$ pc for NGC 7762. We obtained similar mean parallax values for NGC 189 and NGC 7762 compared to \citet{Cantat-Gaudin_2020} and \citet{Cantat-Gaudin-Anders_2020}. NGC 1758 was not examined by those studies. 

\begin{table}
\setlength{\tabcolsep}{3pt}
\renewcommand{\arraystretch}{1.2}
  \centering
  \caption{The median astrometric parameters of the three open clusters. Proper motion components ($\langle\mu_{\alpha}\cos\delta\rangle$, $\langle\mu_{\delta}\rangle$), trigonometric parallaxes ($\langle \varpi\rangle $), and the distances ($d_{\varpi}$) calculated from the trigonometric parallaxes of the cluster's most probable member stars ($P\geq 0.50$).}
    \begin{tabular}{lcccc}
\hline
    Cluster & $\langle\mu_{\alpha}\cos\delta\rangle$ &   $\langle\mu_{\delta}\rangle$  & $\langle \varpi\rangle $ & $d_{\varpi}$\\ 
            & (mas yr$^{-1}$) & (mas yr$^{-1}$)  & (mas) & (pc)\\
\hline
NGC ~~189 & +0.352$\pm$0.032 & -3.412$\pm$0.038 & 0.778$\pm$0.051 &  1285$\pm$84 \\
NGC 1758  & +3.141$\pm$0.035 & -3.507$\pm$0.025 & 1.150$\pm$0.076 & ~~870$\pm$57 \\
NGC 7762  & +1.489$\pm$0.023 & +3.962$\pm$0.024 & 1.033$\pm$0.025 & ~~968$\pm$23 \\
\hline
    \end{tabular}%
    \label{tab:proper_motion}%
\end{table}%

\section{Astrophysical Parameters}
\label{sec:parameters}
In this section, we sum up the methods followed to determine the fundamental parameters of the open clusters NGC 189, NGC 1758, and NGC 7762. We followed the processes outlined in \cite{Yontan_2023}. Dense stellar regions, different membership selection criteria, high/differential extinction towards the cluster regions, and different analyses procedures can influence the determination of the fundamental astrophysical parameters of clusters and result in degeneracies between reddening and age parameters \citep{Anders_2004, King_2005}. Such a situation is the cause of large discrepancies across the estimated parameters performed by various researchers for the same clusters. To avoid possible degeneracy issues and achieve accurate calculations, the color excesses and metallicities of the clusters were derived using $(U-B)\times (B-V)$ two-color diagrams (TCDs) separately, while we obtained distance moduli and ages individually by fitting theoretical models on CMDs. 

\subsection{Color Excess}
\label{sec:color_excess}

\begin{figure}
\centering
\includegraphics[scale=0.7, angle=0]{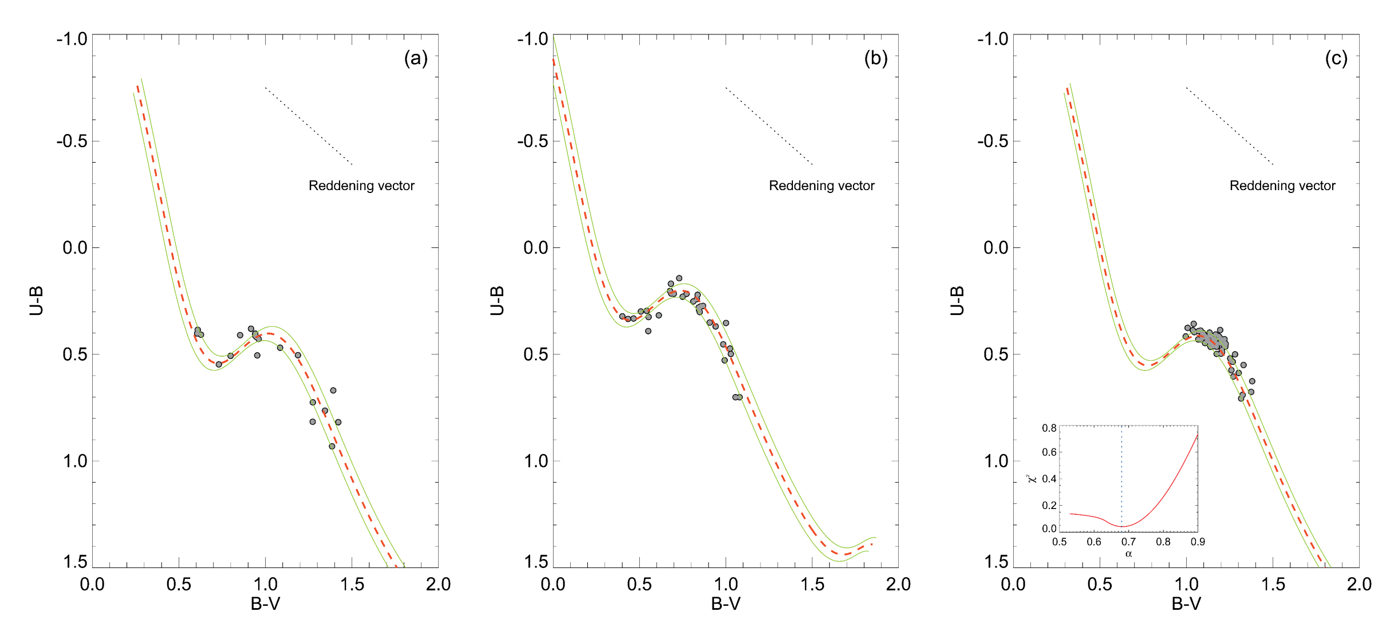}
\caption{$(U-B)\times (B-V)$ diagrams with the best-fitted ZAMS for NGC 189 (a), NGC 1758 (b) and NGC 7762 (c). Green-filled circles represent the main-sequence members with the probability $P\geq 0.5$. Red dashed and green solid curves show the reddened ZAMS of \citet{Sung_2013} and $\pm1\sigma$ standard deviations, respectively. In the inner panel of NGC 7762 (panel c),  $\chi^2$ values versus the estimated reddening slope are presented. The dashed blue line is the best reddening slope based on $\chi^2_{\rm min}$. 
\label{fig:tcds}} 
\end{figure}

Estimation of reddening was performed using $(U-B)\times (B-V)$ TCDs of the clusters. We selected the main-sequence member stars with probability $P\geq 0.5$ and within the magnitude ranges $13\leq V \leq 19$ for NGC 189, $10\leq V \leq 17$ mag for NGC 1758 and $14\leq V \leq 17$ mag for NGC 7762 in order to derive the $E(U-B)$ and $E(B-V)$ color excess values. We plotted $(U-B)\times (B-V)$ TCDs using these member stars and fitted the ZAMS of \citet{Sung_2013} to these distributions. The ZAMS was fitted by considering varied values of the reddening slope $\alpha=E(U-B)/E(B-V)$, applying a $\chi^2$ minimization method. The best result of minimum $\chi^2$ equates to the reddening slope being $\alpha=0.72$ as well as the color excess $E(B-V)$ being $0.590\pm 0.023$ mag for NGC 189 and $0.310\pm 0.022$ mag for NGC 1758. In the case of NGC 7762 the best reddening slope and color excess values are $\alpha=0.68$ and $E(B-V)=0.640\pm 0.017$ mag, respectively. TCDs with the best solutions are presented in Fig.~\ref{fig:tcds}. Color excess errors are $\pm 1\sigma$ deviations. Our estimated color excess $E(B-V)$ for NGC 189 is close to the values found by \citet{Kharchenko_2013}, \citet{Joshi_2016} and \citet{Dias_2021}. For NGC 1758, the estimated $E(B-V)$ is compatible with the studies of \citet{Kharchenko_2005} and \citet{Sampedro_2017}. For NGC 7762 the current estimate is in good agreement with results given by \citet{Kharchenko_2013}, \citet{Maciejewski_2007} and \citet{Cantat-Gaudin_2020}. See Table~\ref{tab:literature} for a detailed comparison.

\subsection{Photometric Metallicities}
\label{sec:metallicities}

\begin{figure*}
\centering
\includegraphics[scale=0.7, angle=0]{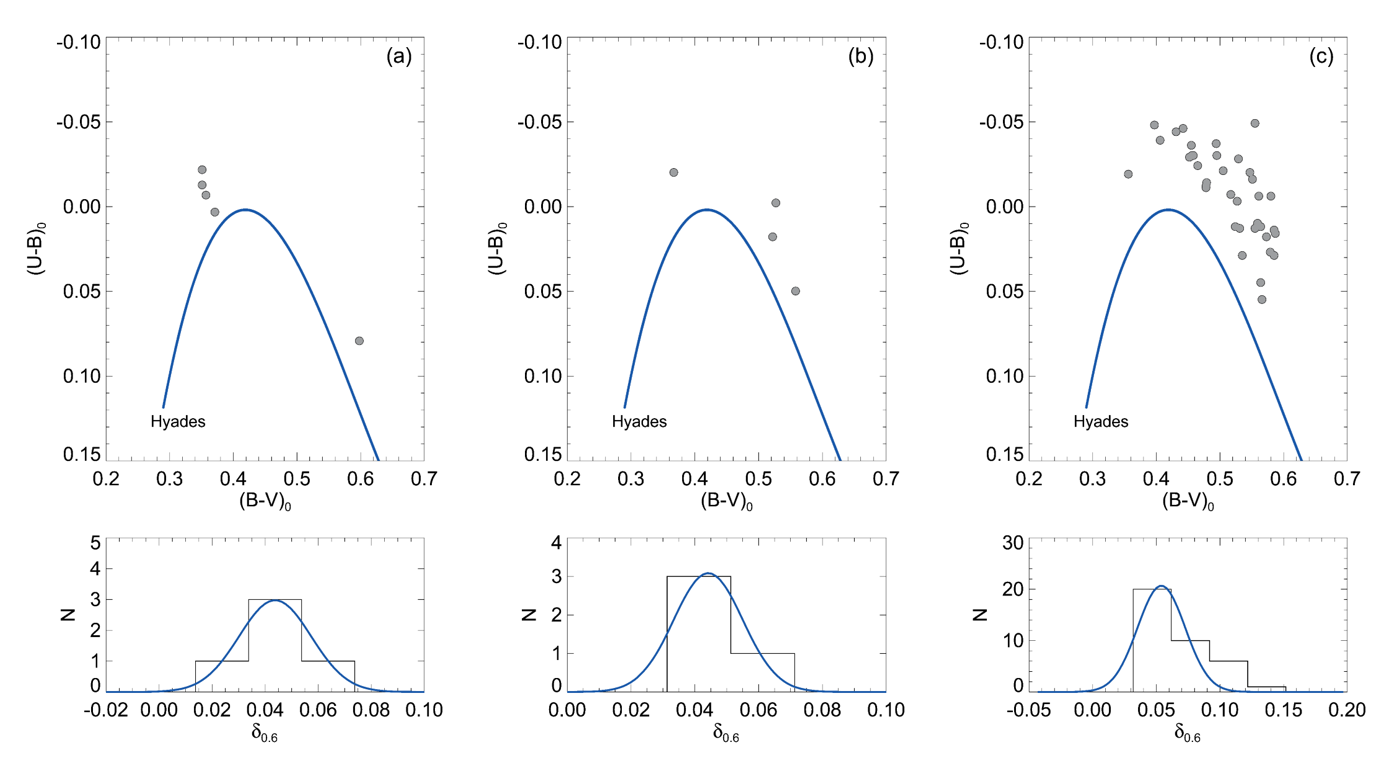}
\caption{$(U-B)_0\times(B-V)_0$ diagrams (upper panels) and histograms for the normalized $\delta_{0.6}$ (lower panels) for five stars (NGC 189, panel a), four stars (NGC 1758, panel b), and 37 stars (NGC 7762, panel c) F-G type main-sequence stars with probability $P\geq 0.5$. The solid blue lines in the upper panels represent the main sequence of Hyades, whereas Gaussian fits in the lower panels.
\label{fig:hyades}} 
\end {figure*}

As with reddening, it is important to know the metallicity before attempting to derive the age and distance of a cluster. Given a lack of spectroscopic observations, $(U-B)\times (B-V)$ TCDs can be used to estimate photometric metallicity \citep{Karaali_2003b, Karaali_2003a, Karaali_2011}. In the present study, we used the method of \citet{Karaali_2011} to determine the photometric metallicity of each cluster. In this method, F and G-type main-sequence stars falling into the $0.3\leq (B-V)_0\leq0.6$ mag color range \citep{Eker_2018, Eker_2020} and their UV-excesses are used. Having calculated de-reddened $(B-V)_0$ and $(U-B)_0$ color indices of the most probable member stars ($P\geq 0.5$) and considering the $E(U-B)$, $E(B-V)$ color excesses derived in the study (see Section~\ref{sec:color_excess}), we made a selection of F-G type main-sequence stars by limiting their color range within $0.3\leq (B-V)_0\leq0.6$ mag. This resulted in the selection of five stars for NGC 189, four stars for NGC 1758 and 37 stars for NGC 7762. Next, the construction of $(U-B)_0\times(B-V)_0$ diagrams (Figure~\ref{fig:hyades} upper panels) was made for these selected stars and compared against that of the Hyades main sequence. This allowed the determination of the differences between the $(U-B)_0$ color indices of the observed cluster members and the Hyades stars which have the same $(B-V)_0$ colors. The calculated $(U-B)_0$ differences are defined as the UV-excess ($\delta$) via the equation $\delta =(U-B)_{\rm 0,H}-(U-B)_{\rm 0,S}$, where H and S are Hyades and cluster stars, respectively. The calculated UV-excesses were calibrated to the normalised $\delta_{0.6}$ values according to $(B-V)_0=0.6$ mag. We plotted histograms of the normalised $\delta_{0.6}$ values and applied Gaussian fits to these distributions to estimate the mean $\delta_{0.6}$ values (Figure~\ref{fig:hyades} lower panels). Considering the peaks of the fitted Gaussian functions, the photometric metallicities of the three open clusters were estimated using the following equation from \citet{Karaali_2011}:
\begin{eqnarray}
{\rm [Fe/H]}=-14.316(1.919)\delta_{0.6}^2-3.557(0.285)\delta_{0.6}+0.105(0.039).
\end{eqnarray}

Fig.~\ref{fig:hyades} presents the TCDs and the distributions of the normalised $\delta_{0.6}$ UV excesses for three clusters. The calculated mean $\delta_{0.6}$ values of NGC 189, NGC 1758, and NGC 7762 are $0.044 \pm 0.014$ mag, $0.044 \pm 0.011$ mag, and $0.053 \pm 0.019$ mag respectively. Thus, the estimated photometric metallicities are [Fe/H]=$-0.08\pm 0.03$ dex for NGC 189, [Fe/H]=$-0.08\pm 0.03$ dex for NGC 1758 and [Fe/H]=$-0.12\pm 0.02$ dex for NGC 7762. Moreover, estimated photometric metallicities ([Fe/H]) were converted to the mass fraction $z$ for choosing isochrones used in age determination. To do this, we took into account the expressions presented from Bovy\footnote{https://github.com/jobovy/isodist/blob/master/isodist/Isochrone.py} which are improved for {\sc parsec} isochrones \citep{Bressan_2012}. The related expressions are given as follows: 

\begin{equation}
z_{\rm x}={10^{{\rm [Fe/H]}+\log \left(\frac{z_{\odot}}{1-0.248-2.78\times z_{\odot}}\right)}}
\end{equation}      
and
\begin{equation}
z=\frac{(z_{\rm x}-0.2485\times z_{\rm x})}{(2.78\times z_{\rm x}+1)}.
\end{equation} 
Where $z_{\rm x}$ and $z_{\odot}$ are intermediate values where solar metallicity $z_{\odot}$ was adopted as 0.0152 \citep{Bressan_2012}. We obtained $z=0.013$ for NGC 189 and NGC 1758 as well as  $z=0.012$ for NGC 7762.

\subsection{Distance and Age}
\label{sec:distance}

In the present study, the age and distance moduli of the studied clusters were estimated simultaneously by fitting {\sc parsec} isochrones \citep{Bressan_2012} to the observed CMDs. We plotted $V\times (U-B)$, $V\times (B-V)$ and $G\times (G_{\rm BP}-G_{\rm RP})$ CMDs and fitted selected isochrones via visual inspection, taking into account the main-sequence, turn-off, and giant stars with membership probabilities $P\geq 0.5$. For each cluster, selection of {\sc parsec} isochrones\footnote{http://stev.oapd.inaf.it/cgi-bin/cmd} was made according to the mass fraction $z$ calculated from photometric metallicities earlier in this study (Section~\ref{sec:metallicities}). For the $U\!BV$ data-driven isochrone fitting process, the isochrones were reddened in regard to the color excesses $E(U-B)$ and $E(B-V)$ as determined in this study, while the {\it Gaia} DR3 reddened procedure employed the equation $E(G_{\rm BP}-G_{\rm RP})= 1.41\times E(B-V)$ as given by \citet{Sun_2021}. Errors for distance moduli and distances were obtained from the expression presented by \citet{Carraro_2017}. Uncertainties in age were estimated by considering the scatter of the observed data representing the main sequence, turn off and giant member stars. We, therefore, selected low and high-age isochrones whose values were close to the cluster's estimated age, taking into account this scatter. $V\times (U-B)$, $V\times (B-V)$ and $G\times (G_{\rm BP}-G_{\rm RP})$ CMDs with the best fitting isochrones are given as Fig.~\ref{fig:figure_age}. Estimated distance moduli and ages for each cluster are as follows:

\begin{figure*}
\centering
\includegraphics[scale=1, angle=0]{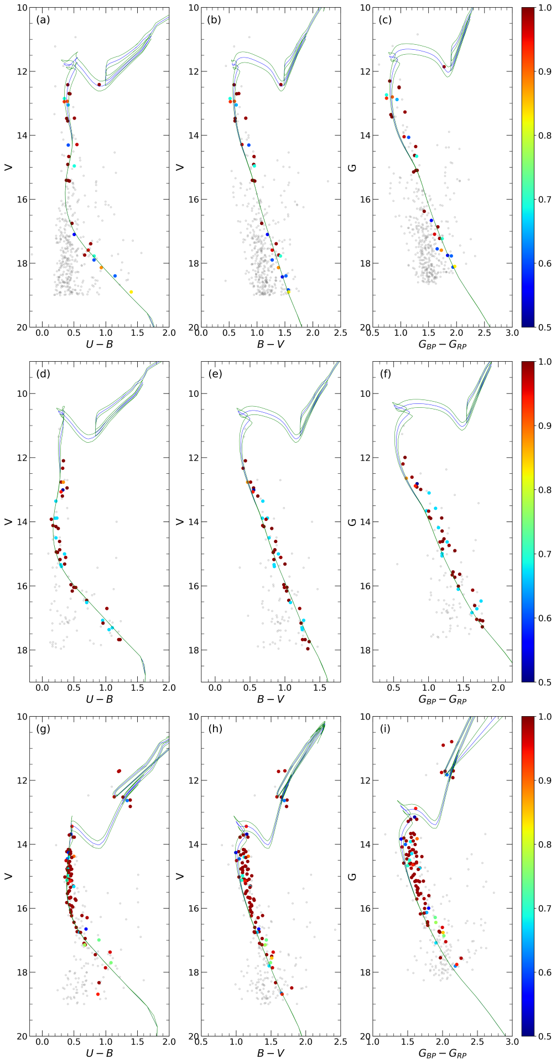}
\caption{Color-magnitude diagrams for the studied clusters NGC 189 (panels a, b and c), NGC 1758 (panels d, e and f) and NGC 7762 (panels g, h and i). Different color correspond to the membership probabilities of the most probable cluster members. The scales of membership probabilities are represented on color bars to the right of the figure. Grey colored dots identify the stars with probabilities $P<0.5$. The best fitting {\sc parsec} isochrones and their errors are shown as the blue lines and green dashed lines, respectively. Superimposed isochrone ages match to 500 Myr for NGC 189, 650 Myr for NGC 1758 and 2000 Myr for NGC 7762.
\label{fig:figure_age} }
\end {figure*}

\begin{itemize}
\item{{\bf NGC 189}: We over-plotted the different isochrones of $\log({\rm age})$ = 8.65, 8.70 and 8.74 with $z=0.013$ on all the CMDs for the cluster NGC 189, as shown in Fig.~\ref{fig:figure_age} (panels a, b and c). The age of NGC 189 is estimated from the well-fitting isochrone $\log({\rm age}) = 8.70$ log-years to the most probable main-sequence and giant stars, and corresponds to the age $t=500\pm 50$ Myr. The derived age is in good agreement with the results of \citet{Kharchenko_2013}, \citet{Joshi_2016} and \citet{Loktin_2017}. We derived the distance modulus of the NGC 189 as $\mu_{\rm V}=12.227\pm 0.094$ mag, which leads to the isochrones-based distance for the cluster being $d_{\rm iso}=1201\pm 53$ pc. This distance value is close to most of the values given by many authors \citep[etc.][see Table~\ref{tab:literature} for detailed comparison] {Cantat-Gaudin_2018, Cantat-Gaudin_2020}. The isochrone distance estimated for the cluster is compatible with the astrometric distance value ($d_{\varpi}=1285\pm 84$ pc) calculated from {\it Gaia} DR3 trigonometric parallaxes of member stars. We estimated the Galactocentric distance as $R_{\rm gc}=8.69$ kpc by considering the Sun's distance to the Galactic centre as 8 kpc \citep{Majewski_1993}. The Galactocentric coordinates were derived as  $(X, Y, Z)_{\odot}=(-627, 1024, -36)$ pc, respectively. These are compatible with the values of \citet{Cantat-Gaudin_2018}.}

\item{{\bf NGC 1758}: We over-plotted the isochrones $\log({\rm age}) = 8.78, 8.81$ and 8.84 with $z =0.013$ to the observed CMDs as presented in Fig.~\ref{fig:figure_age} (panels d, e and f). The best-fit isochrone corresponded to the age of the NGC 1758 being $\log({\rm age}) = 8.70$ or a value of $t=650\pm 50$ Myr. This age estimate is about 100 Myr older than the estimated ages presented by \citet{Bossini_2019} and \citet{Dias_2021}. We estimated the distance modulus of the NGC 1758 as $\mu_{\rm V}=10.736\pm 0.078$ mag. This corresponds to the isochrone-based distance for the cluster being $d_{\rm iso}=902\pm 33$ pc. This result is close to the values given by \citet{Liu_2019} and \citet{Bossini_2019}. It is also compatible with the results given by \citet{Cantat-Gaudin_2018}, \citet{Cantat-Gaudin-Anders_2020}, and \citet{Dias_2021} (see Table~\ref{tab:literature} for a detailed comparison). The isochrone distance estimated for the NGC 1758 is compatible with the astrometric distance value of $d_{\varpi}=870\pm 57$ pc. The Galactocentric distance and Galactocentric coordinates were obtained as $R_{\rm gc}=8.89$ kpc and $(X, Y, Z)_{\odot}=(-887, 13, -164)$ pc. These are in good agreement with the values of \citet{Cantat-Gaudin_2018}.}

\item{{\bf NGC 7762}: Isochrones of ages $\log({\rm age}) = 9.25, 9.30$ and 9.35 were over-plotted with $z=0.012$ to the $U\!BV$  and {\it Gaia} data-based CMDs, as shown in Fig.~\ref{fig:figure_age} (panels g, h and i). Considering the most probable main sequence, turn-off and giant cluster members, we concluded that the best fitting isochrone gives the age of the NGC 7762 as $\log({\rm age}) = 9.30$, which corresponds to $t=2000\pm 200$ Myr. The estimated age of the cluster is in good agreement with the values of \citet{Maciejewski_2007}, \citet{Joshi_2016} and \citet{Cantat-Gaudin_2020}. We obtained the distance modulus of the NGC 7762 as $\mu_{\rm V}=11.781\pm 0.072$ mag, which gives the isochrone-based distance of the cluster as $d_{\rm iso}=911\pm 31$ pc. This value is compatible with the values given in many studies which are based on {\it Gaia} data \citep[][see Table~\ref{tab:literature} for a detailed comparison] {Soubiran_2018, Cantat-Gaudin_2020, Cantat-Gaudin-Anders_2020}. The isochrone distance is close to the distance value ($d_{\varpi}=968\pm 23$ pc) calculated from {\it Gaia} DR3 trigonometric parallaxes in this study. The Galactocentric distance and Galactocentric coordinates were obtained as $R_{\rm gc}=8.45$ kpc and $(X, Y, Z)_{\odot}=(-414, 806, 93)$ pc. These are compatible with the corresponding estimates of \citet{Cantat-Gaudin_2018}}.

\end{itemize}
\noindent

\section{Kinematics and Galactic orbit parameters}
\label{sec:kinematics}
We estimated the space velocity components and Galactic orbit parameters of NGC 189, NGC 1758 and NGC 7762 using the Python package {\sc galpy} \cite{Bovy_2015}\footnote{See also https://galpy.readthedocs.io/en/v1.5.0/}. To perform integrations of the Galactic orbits for these clusters, we adopted the default {\sc galpy} potential of the Milky Way that assumes an axisymmetric potential for the Galaxy ({\sc MWPotential2014}, \citealt{Bovy_2015}). The {\sc MilkywayPotential2014} code is described by a three-component model consisting of the bulge, disc, and halo potentials. The bulge is described as a spherical power-law density profile according to the expression of \cite{Bovy_2015}:  

\begin{equation}
\rho (r) = A \left( \frac{r_{\rm 1}}{r} \right) ^{\alpha} \exp \left[-\left(\frac{r}{r_{\rm c}}\right)^2 \right] \label{eq:rho}
\end{equation} 
where $r_{\rm 1}$ and $r_{\rm c}$ are the present reference radius and cut-off radius, respectively. $A$ and $\alpha$ denote the amplitude that is applied to the potential in mass density units and inner power, respectively. The Galactic disc component that we used is expressed by \citet{Miyamoto_1975} as:

\begin{equation}
\Phi_{\rm disc} (R_{\rm gc}, Z) = - \frac{G M_{\rm d}}{\sqrt{R_{\rm gc}^2 + \left(a_{\rm d} + \sqrt{Z^2 + b_{\rm d}^2 } \right)^2}} \label{eq:disc}
\end{equation}
where $R_{\rm gc}$ and $Z$ are the distances from the Galactic centre and Galactic plane, respectively. $G$ is the universal gravitational constant, and $M_{\rm d}$ the mass of the Galactic disc, $a_{\rm d}$ the scale length of the disc and $b_{\rm d}$ the height of the disc. The halo component is assumed as a spherically symmetric spatial distribution of the dark matter density. The expression of \citet{Navarro_1996} was adopted to model the halo as:

\begin{equation}
\Phi _{\rm halo} (r) = - \frac{G M_{\rm s}}{R_{\rm gc}} \ln \left(1+\frac{R_{\rm gc}}{r_{\rm s}}\right) \label{eq:halo}
\end{equation} 
where $M_{\rm s}$ denotes the mass of the dark matter halo of the Milky Way and $r_{\rm s}$ represents its radius. The galactocentric distance and the circular velocity of the Sun were taken as $R_{\rm gc}=8$ kpc and $V_{\rm rot}=220$ km s$^{-1}$, respectively \citep{Bovy_2012, Bovy_2015}. The Sun's distance from the Galactic plane was taken as 27$\pm$4 pc \citep{Chen_2001}.

The input parameters for the orbital integration of the three clusters are listed in Table~\ref{tab:Final_table}, namely the equatorial coordinates ($\alpha,~\delta$) estimated in the study (Section~\ref{sec:structural}), proper motion components ($\mu_{\alpha}\cos \delta, \mu_{\delta}$) estimated in Section~\ref{sec:cmds}, distance ($d_{\rm iso}$) from the Section~\ref{sec:distance}, and radial velocity ($V_{\gamma}$). The radial velocities of the clusters were derived using the data available in the {\it Gaia} DR3 catalogue. To calculate the mean radial velocities (and their uncertainties) of the clusters, we selected the stars with membership probabilities $P\geq 0.9$ and used the weighted average of their radial velocities taken from {\it Gaia} DR3 data \citep[for equations see ][]{Soubiran_2018}. The mean radial velocities were calculated as $-29.60 \pm 0.32$ km s$^{-1}$ for NGC 189 from five member stars, as $+6.80 \pm 2.90$ km s$^{-1}$ for NGC 1758 from eight members, and $-46.61 \pm 0.10$ km s$^{-1}$ for NGC 7762 from 31 members. The literature mean radial velocity estimates for NGC 189 are $-28.47 \pm 0.75$ \citep{Soubiran_2018} and $-29.391 \pm 0.534$ km s$^{-1}$ \citep{Dias_2021}, while for NGC 1758 the literature value is $17.840 \pm 9.887$ km s$^{-1}$ \citep{Dias_2021}. The values for NGC 7762 are $-45.7 \pm 0.7$ km s$^{-1}$ \citep{Casamiquela_2016}, $-45.40 \pm 0.14$ km s$^{-1}$ \citep{Soubiran_2018} and $-45.438 \pm 0.492$ km s$^{-1}$  \citep{Dias_2021} (Table~~\ref{tab:literature}). The mean radial velocities derived in this study for NGC 189 and NGC 7762 are within 1-2 km s$^{-1}$ of the values given by these different researchers. In the case of NGC 1758, there is a difference between the estimated and literature mean radial velocity value. We used radial velocities of eight stars with membership probabilities $P\geq 0.9$ for NGC 1758, whereas in the study of \citet{Dias_2021} the mean radial velocity of the cluster was estimated from two members. We concluded that the reason for the discrepancies between these two studies is the number and membership probabilities of the stars used in calculations.

Kinematic and dynamic calculations were analyzed with 1 Myr steps over a 3 Gyr integration time. Estimates for the apogalactic distance $R_{\rm a}$, perigalactic  distance $R_{\rm p}$, eccentricity $e$, the maximum vertical distance from Galactic plane $Z_{\rm max}$, Galactic space velocity components ($U$, $V$, $W$), and orbital period $P_{\rm orb}$ were derived and listed in Table~\ref{tab:Final_table}. The  space velocity components $(U, V, W)$ were estimated as $(14.83 \pm 0.37, -26.31 \pm 0.12, -18.59 \pm 0.78)$ km s$^{-1}$ for NGC 189, $(-7.38 \pm 2.91, -19.91 \pm 0.78, 0.71 \pm 0.28)$ km s$^{-1}$ for NGC 1758 and $(12.73\pm 0.50, -47.32 \pm 0.23, 10.27 \pm 0.71)$ km s$^{-1}$ for NGC 7762. To apply the local standard of rest (LSR), we employed the velocity components of \citet{Coskunoglu_2011} as ($U, V, W$) = $(8.83 \pm 0.24$, $14.19 \pm 0.34$, $6.57 \pm 0.21)$ km s$^{-1}$. Hence, we derived transformed parameters $(U, V, W)_{\rm LSR}$ as $(23.66 \pm 0.44, -12.12 \pm 0.36, -12.02 \pm 0.80)$ km s$^{-1}$ for NGC 189, $(1.46 \pm 2.92, -5.72 \pm 0.85, 7.28 \pm 0.35)$ km s$^{-1}$ for NGC 1758 and $(21.56 \pm 0.56, -33.13 \pm 0.41, 16.84 \pm 0.74)$ km s$^{-1}$ for NGC 7762. Moreover, the total space velocities ($S_{\rm LSR}$) of NGC 189 were calculated to be $29.17 \pm 0.98$ km s$^{-1}$, $9.37 \pm 3.06$ km s$^{-1}$ for NGC 1758 and $42.96 \pm 1.02$ km s$^{-1}$ for NGC 1798 (for all estimated parameters see Table~\ref{tab:Final_table}). To obtain the birth radii of three clusters, we utilized the orbital integration in the past epochs, using time bins equal to estimated cluster ages. 

\begin{figure*}
\centering
\includegraphics[width=\textwidth]{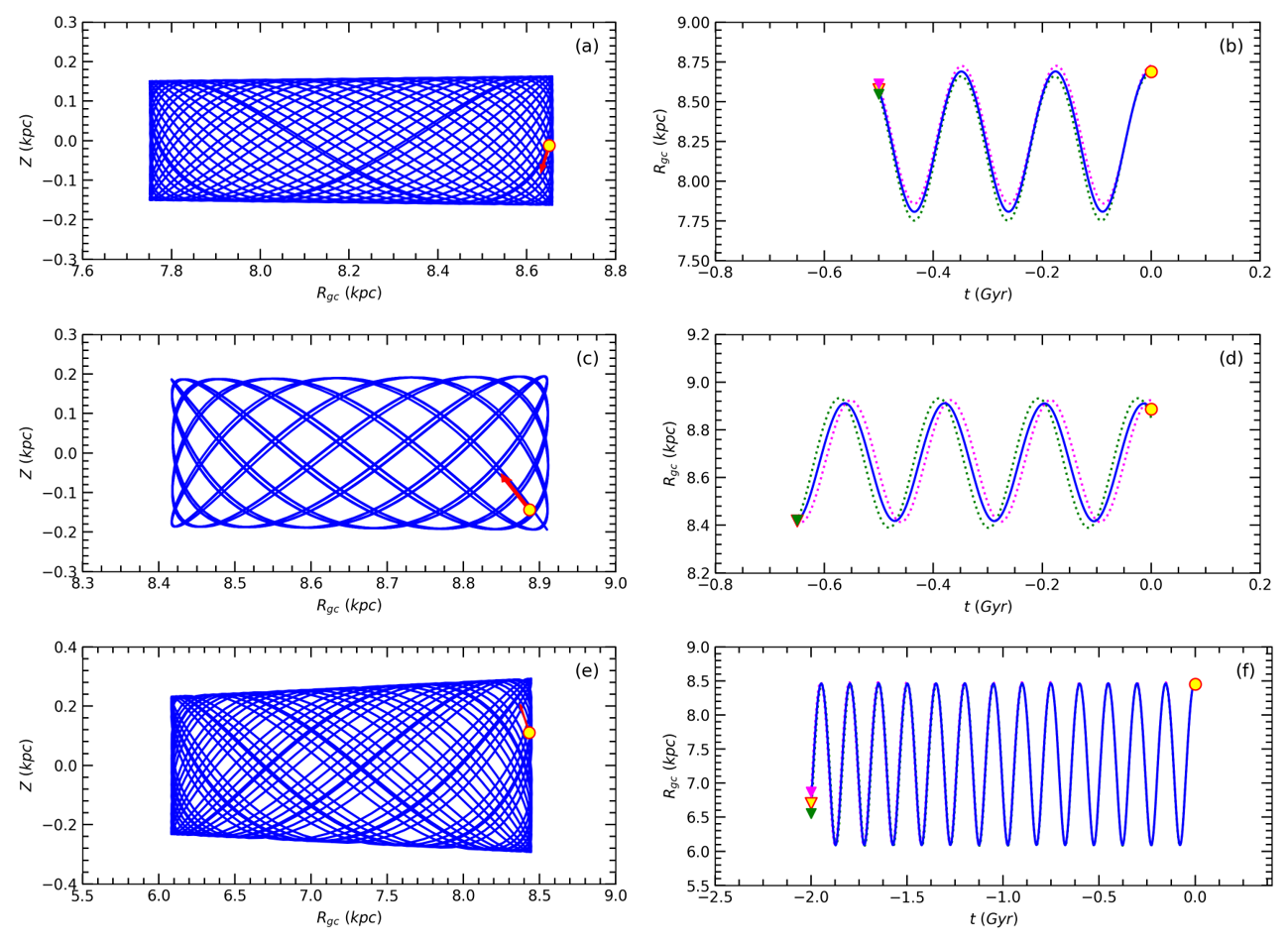}
\caption{\label{fig:galactic_orbits}
The Galactic orbits and birth radii of NGC 189 (a, b), NGC 1758 (c, d) and NGC 7762 (e, f) in the $Z \times R_{\rm gc}$ and $R_{\rm gc} \times t$ planes. The filled yellow circles and triangles show the present-day and birth positions, respectively. Red arrows are the motion vectors of open clusters. The green and pink dotted lines
show the orbit when errors in input parameters are considered, while the green and pink filled triangles represent the birth locations of the open cluster based on the lower and upper error estimates.} 
\end {figure*}

The Galactic orbits of the three clusters are represented in Fig.~\ref{fig:galactic_orbits}. The left panels of this figure show the `side views' of the cluster motions (panel a for NGC 189; panel c for NGC 1758 and panel e for NGC 7762) in respect of distance from the Galactic center and the Galactic plane. According to the analyses, both NGC 189 and NGC 1758 follow a `boxy' pattern with almost zero eccentricities, whereas the orbit of the NGC 7762 differs from a `boxy' pattern given its larger eccentricity. We investigated the orbital parameters to understand how uncertainties in the input parameters ($\mu_{\alpha}\cos \delta, \mu_{\delta}$, $d_{\rm iso}$, and $V_{\gamma}$) influence the orbit integration results and estimated birth-radii of the clusters. In the right panel of the Fig.~\ref{fig:galactic_orbits} the motion in the Galactic disc and the effect of uncertainties in the input parameters are presented in terms of time for the three clusters (panel b for NGC 189; panel d for NGC 1758, and panel f for NGC 7762). It can be seen from the figure that the range in possible birth radii, between the extreme estimates, do not exceed 0.4 kpc for any of the three clusters. 

These estimates for the birth-radii indicate that both NGC 189 and NGC 1758 were formed outside the solar vicinity at 8.58 and 8.42 kpc, while NGC 7762 was formed inside the solar vicinity with a birth radius of 6.70 kpc. According to the perigalactic and apogalactic distances of the three clusters, it can be concluded that the orbit of NGC 1758 is completely outside the solar circle (Fig.~\ref{fig:galactic_orbits}e), while NGC 189 and NGC 7762 cross the solar circle during their orbits (Fig.~\ref{fig:galactic_orbits}a and c, respectively). All three clusters rise above the plane up to $Z_{\rm max}=310$ pc. This indicates that NGC 189, NGC 1758 and NGC 7762 belong to the thin-disc component of the Milky Way \citep{Bilir_2006b, Bilir_2006c, Bilir_2008}.  

\section{Summary \& Conclusions}
\label{sec:summary}

We presented photometric, astrometric, and kinematic studies of the open clusters NGC 189, NGC 1758 and NGC 7762 using CCD $U\!BV$ and {\it Gaia} DR3 data. We identified 32, 57 and 106 member stars (with estimated membership probabilities $P\geq 0.5$) in the regions of NGC 189, NGC 1758, and NGC 7762, respectively. These members were employed in the estimation of the main astrophysical parameters of the clusters. We used $U\!BV$ data-based TCDs, deriving color excesses and photometric metallicities separately to achieve reliable age and distance values. We also investigated the kinematical properties of the clusters. The results of these analyses are listed in Table~\ref{tab:Final_table}. In literature studies, parameters were derived simultaneously through statistical solutions for the three clusters. This can cause degeneracies between the parameters (reddening, age and distance) and large uncertainties in the measured values. We therefore believe that the approach described in this paper leads to more reliable estimates.

The main findings of the current study are summarized as follows:   

\begin{table}
\renewcommand{\arraystretch}{1.5}
  \centering
  \caption{Fundamental parameters of NGC 189, NGC 1758 and NGC 7762.}
  {\small
        \begin{tabular}{lrrr}
\hline
Parameter & NGC 189 & NGC 1758 & NGC 7762 \\
\hline
($\alpha,~\delta)_{\rm J2000}$ (Sexagesimal)& 00:39:27.89, $+$61:06:46.35& 05:04:42.00, $+23$:48:46.90 & 23:49:53.00, $+68$:02:06.00\\
($l, b)_{\rm J2000}$ (Decimal)              & 121.4858, $-01.7252$      & 179.1584, $-10.4597$  & 117.2060, $+05.8501$  \\
$f_{0}$ (stars arcmin$^{-2}$)               & $4.86 \pm 0.41$           & $2.97 \pm 0.17$       & $4.91 \pm 0.24$       \\
$f_{\rm bg}$ (stars arcmin$^{-2}$)          & $12.38 \pm 0.08$          & $3.14 \pm 0.08$       & $4.92 \pm 0.13$       \\
$r_{\rm c}$ (arcmin)                        & $0.81 \pm 0.12$           & $2.13 \pm 0.28$       & $4.39 \pm 0.46$       \\
$r_{\rm lim}$ (arcmin)                      & 4                         & 7                     & 15                    \\
$r$ (pc)                                    & 1.40                      & 1.98                  & 3.97                  \\
Cluster members ($P\geq0.5$)                & 32                        & 57                    & 106                    \\
$\mu_{\alpha}\cos \delta$ (mas yr$^{-1}$)   & $0.352 \pm 0.032$        & $3.141 \pm 0.035$     & $1.489 \pm 0.023$     \\
$\mu_{\delta}$ (mas yr$^{-1}$)              & $-3.412 \pm 0.038$        &$-3.507 \pm 0.025$     & $3.962 \pm 0.024$     \\
$\varpi$ (mas)                              & $0.778 \pm 0.051$         & $1.150 \pm 0.076$     & $1.033 \pm 0.025$       \\
$d_{\varpi}$ (pc)                           & $1285\pm 84$              & $870\pm 57$           & $968\pm 23$           \\
$E(B-V)$ (mag)                              & $0.590 \pm 0.023$         & $0.310 \pm 0.022$     & $0.640 \pm 0.017$     \\
$E(U-B)$ (mag)                              & $0.425 \pm 0.017$         & $0.223 \pm 0.016$     & $0.435 \pm 0.012$     \\
$A_{\rm V}$ (mag)                           & $1.829 \pm 0.071$         & $0.961 \pm 0.068$     & $1.984 \pm 0.053$     \\
$E(G_{\rm BP}-G_{\rm RP})$ (mag)            & $0.832 \pm 0.032$         & $0.437 \pm 0.031$     & $0.902 \pm 0.024$     \\
$A_{\rm G}$ (mag)                           & $1.548 \pm 0.026$         & $0.814 \pm 0.047$     & $1.680 \pm 0.039$     \\
$[{\rm Fe/H}]$ (dex)                        & $-0.08 \pm 0.03$          & $-0.08 \pm 0.03$      & $-0.12 \pm 0.02$      \\
Age (Myr)                                   & $500 \pm 50$              & $650 \pm 50$          & $2000 \pm 200$        \\
Distance modulus (mag)                      & $12.227 \pm 0.094$        & $10.736\pm 0.078$     & $11.781 \pm 0.072$    \\
Isochrone distance (pc)                     & $1201 \pm 53$             & $902 \pm 33$          & $911 \pm 31$          \\
$(X, Y, Z)_{\odot}$ (pc)                    & ($-627$, 1024, -36)       & ($-887$, 13, -164)    & ($-414$, 806, 93)     \\
$R_{\rm gc}$ (kpc)                          & 8.69                      & 8.89                  & 8.45                  \\
$V_{\gamma}$ (km/s)                         & $-29.60 \pm 0.32$         & $6.80 \pm 2.90$       & $-46.61 \pm 0.10$     \\
$U_{\rm LSR}$ (km/s)                        & $23.66 \pm 0.44$          & $1.46 \pm 2.92$       & $21.56 \pm 0.56$      \\
$V_{\rm LSR}$ (km/s)                        & $-12.12 \pm 0.36$         & $-5.72 \pm 0.85$      & $-33.13 \pm 0.41$     \\
$W_{\rm LSR}$ (km/s)                        & $-12.02 \pm 0.80$         & $7.28 \pm 0.35$       & $16.84 \pm 0.74$      \\
$S_{_{\rm LSR}}$ (km/s)                     & $29.17 \pm 0.98$          & $9.37 \pm 3.06$       & $42.96 \pm 1.02$      \\
$R_{\rm a}$ (pc)                            & $8690 \pm 36$             & $8912 \pm 22$         & $8466 \pm 17$         \\
$R_{\rm p}$ (pc)                            & $7808 \pm 56$             & $8418 \pm 28$         & $6091 \pm 7$          \\
$z_{\rm max}$ (pc)                          & $177 \pm 14$              & $194\pm 3$            & $307 \pm 16$          \\
$e$                                         & $0.053 \pm 0.002$         & $0.029\pm 0.003$      & $0.163 \pm 0.001$     \\
$P_{\rm orb}$ (Myr)                         & $231 \pm 2$               & $244 \pm 1$           & $203 \pm 1$           \\
Birthplace (kpc)                            & 8.58                      & 8.42                  & 6.70                  \\
\hline
        \end{tabular}%
    } 
    \label{tab:Final_table}%
\end{table}%

\begin{enumerate}[1.]

\item We investigated the structural properties of the clusters. We used the central coordinates determined in the study during RDP analyses. The centers of three clusters are: 
\begin{itemize}
    \item ($\alpha=00^{\rm h} 39^{\rm m} 27^{\rm s}\!\!.89, \delta=+61^{\rm o} 06^{'} 46^{''}\!\!.35$) for NGC 189, 
    \item ($\alpha=05^{\rm h} 04^{\rm m} 42^{\rm s}\!\!.00, \delta=+23^{\rm o} 48^{'} 46^{''}\!\!.90$) for NGC 1758, and
    \item ($\alpha=23^{\rm h} 49^{\rm m} 53^{\rm s}\!\!.00, \delta=+68^{\rm o} 02^{'} 06^{''}\!\!.00$) for NGC 7762. 
\end{itemize}
The limiting radii $r_{\rm lim}$ were obtained as $4^{'}$ for NGC 189, $7^{'}$ for NGC 1758, and $15^{'}$ for NGC 7762. Astrometric analyzes showed that NGC 1746 does not exist, but has been misidentified in the past and instead is NGC 1758.

\item We separated cluster members from the field stars by considering $U\!BV$ data-based CMDs as well as the {\it Gaia} DR3 proper motion and trigonometric parallax data. We utilized the {\sc upmask} method to calculate the membership probabilities of stars. Based on the distribution of the most probable member stars ($P\geq 0.5$) on $V\times (B-V)$ CMDs, we fitted ZAMS curves that included the effect of possible binary star contamination on the cluster main sequence. From the location of these selected stars on VPDs, we derived mean proper motion components as:
\begin{itemize}
    \item ($-0.352\pm 0.032, -3.412\pm 0.038$) mas yr$^{-1}$ for NGC 189, 
    \item ($3.141\pm 0.035, -3.507\pm 0.025$) mas yr$^{-1}$ for NGC 1758 and
    \item ($1.489\pm 0.023, 3.962\pm 0.024$) mas yr$^{-1}$ for NGC 7762.
\end{itemize} Utilizing the Gaussian fits on histograms constructed from most probable member stars' {\it Gaia} DR3 trigonometric parallaxes ($\varpi$), we determined mean trigonometric parallaxes as  $0.778 \pm 0.051$ mas for NGC 189, $1.150 \pm 0.076$ mas for NGC 1758 and $1.033 \pm 0.025$ mas for NGC 7762. Using the  equation $d({\rm pc})=1000/\varpi$ (mas), we converted the mean trigonometric values into astrometric distances ($d_{\varpi}$): 
\begin{itemize}
    \item $1285\pm 84$ pc for NGC 189, 
    \item $870\pm 57$ pc for NGC 1758 and 
    \item $968\pm 23$ pc for NGC 7762.  
\end{itemize}

\item Comparison of the $(U-B)\times (B-V)$ diagrams of most probable member stars with the ZAMS \citep{Sung_2013} via $\chi^2$ minimization led to estimates of the reddening slope $\alpha=E(U-B)/E(B-V)$ and color excesses $E(B-V)$ for the clusters. The best fitting solutions corresponded to $\alpha = 0.72$ and $E(B-V) = 0.590 \pm 0.023$ mag for NGC 189, 0.72 and $0.310 \pm 0.022$ mag for NGC 1758 and 0.68 and $0.640 \pm 0.017$ mag for NGC 7762.

\item Estimation of photometric metallicity is based on the calculation of UV-excesses of selected F-G main-sequence stars with membership probabilities $P\geq 0.5$, compared to the  Hyades main-sequence on $(U-B)\times (B-V)$ diagrams. The values of photometric metallicities ([Fe/H]) were calculated as $-0.08 \pm 0.03$ dex for NGC 189, $-0.08 \pm 0.03$ dex for NGC 1758 and $-0.12 \pm 0.02$ dex for NGC 7762. We transformed the [Fe/H] values to mass fractions $z$ to derive the ages and distances of the clusters (in the following step below). These are $z = 0.013$ for NGC 189 and NGC 1758, while $z = 0.012$ for NGC 7762.  

\item Keeping the color excess and metallicity parameters as constants, we derived distance moduli and ages by fitting {\sc parsec} isochrones to the CMDs of the clusters constructed from $U\!BV$ and {\it Gaia} DR3 data. The distance moduli ($\mu_{\rm V}$), isochrone distances ($d_{\rm iso}$), and ages ($t$) were found to be:
\begin{itemize}
    \item $12.227 \pm 0.094$ mag, $1201 \pm 53$ pc,  $500 \pm 50$ Myr for NGC 189, respectively;
    \item $10.736\pm 0.078$  mag, $902 \pm 33$ pc, and $650 \pm 50$ Myr for NGC 1758 and 
    \item $11.781 \pm 0.072$ mag, $911 \pm 31$ pc, and $2000 \pm 200$  Myr for NGC 7762. 
\end{itemize}

\item We calculated the mean radial velocities of the clusters using only the stars with membership probabilities $P\geq 0.9$ and taking into account the radial velocities available in {\it Gaia} DR3 database.  This allowed us to estimate the kinematic properties of the clusters. The values of the mean radial velocities ($V_{\gamma}$) were estimated to be $-29.60 \pm 0.32$  km s$^{-1}$ for NGC 189, $6.80 \pm 2.90$ km s$^{-1}$ for NGC 1758 and $-46.61 \pm 0.10$ km s$^{-1}$ for NGC 7762. 

\item Galactic orbits and their relevant output parameters were derived using the {\sc MWPotential2014} of \citet{Bovy_2015}. We concluded that NGC 189 and NGC 7762 are following `boxy' orbits. Estimates for the birth-radii indicate that both NGC 189 and NGC 1758 were formed outside the solar vicinity at 8.58 and 8.42 kpc, respectively while NGC 7762 was formed inside the solar vicinity with a birth radius of 6.70 kpc.

\end{enumerate}
\section*{Acknowledgments}
This study has been supported in part by the Scientific and Technological Research Council (T\"UB\.ITAK) 122F109.
The observations of this publication were made at the National Astronomical Observatory, San Pedro M{\'a}rtir, Baja California, M{\'e}xico, and the authors wish to thank the staff of the Observatory for their assistance during these observations. This research has made use of the WEBDA database, operated at the Department of Theoretical Physics and Astrophysics of the Masaryk University. We also  made use of NASA's Astrophysics Data System as well as the VizieR and Simbad databases at CDS, Strasbourg, France and data from the European Space Agency (ESA) mission \emph{Gaia}\footnote{https://www.cosmos.esa.int/gaia}, processed by the \emph{Gaia} Data Processing and Analysis Consortium (DPAC)\footnote{https://www.cosmos.esa.int/web/gaia/dpac/consortium}. Funding for DPAC has been provided by national institutions, in particular the institutions participating in the \emph{Gaia} Multilateral Agreement. IRAF was distributed by the National Optical Astronomy Observatory, which was operated by the Association of Universities for Research in Astronomy (AURA) under a cooperative agreement with the National Science Foundation. PyRAF is a product of the Space Telescope Science Institute, which is operated by AURA for NASA. We thank the University of Queensland for the collaboration software. We are grateful to the anonymous referees for their feedback, which improved the paper.
\bibliographystyle{jasr-model5-names}
\biboptions{authoryear}
\bibliography{refs}
\end{document}